\documentclass[10pt,aps,pra,twocolumn,nofootinbib,superscriptaddress,floatfix,noeprint]{revtex4-2}
\usepackage[tbtags]{mathtools}
\usepackage{graphicx}
\graphicspath{{figures}}
\usepackage[protrusion,expansion,tracking,kerning,spacing]{microtype}
\usepackage{physics}
\usepackage[final,pdfencoding=auto,hidelinks,colorlinks=true,
  linkcolor=blue,
  filecolor=blue,
  citecolor = black,
  urlcolor=cyan]{hyperref} 
\usepackage[capitalize]{cleveref}
\usepackage{xstring}
\usepackage{natbib}
\usepackage[upint]{stix2}
\usepackage{siunitx}
\usepackage{xcolor}
\usepackage{mathdots}
\usepackage{aligned-overset}
\usepackage{makecell}
\usepackage{cancel} 
\usepackage{poormanssubfig}
\crefname{subfigure}{Fig.}{Figs.}
\usepackage{orcidlink}

\def\sys{\ensuremath{\mathrm{S}}}
\def\bath{\ensuremath{\mathrm{B}}}
\def\experimental{\ensuremath{\mathrm{exp}}}
\def\inter{\ensuremath{\mathrm{I}}}

\def\lamb{\ensuremath{\Delta_{\mathrm{L}}}}

\newcommand{\hc}{\ensuremath{\mathrm{h.c.}}}

\newcommand{\eu}{\ensuremath{\operatorname{e}}}
\newcommand{\iu}{\ensuremath{\mathrm{i}\mkern1mu}}

\newcommand{\abss}[1]{\abs{#1}^{2}}

\DeclareMathOperator{\sinc}{sinc}

\def\ZZ{\ensuremath{\mathbb{Z}}}

\DeclareFontFamily{U}{wncy}{}
\DeclareFontShape{U}{wncy}{m}{n}{<->wncyr10}{}
\DeclareSymbolFont{mcy}{U}{wncy}{m}{n}
\DeclareMathSymbol{\Sh}{\mathord}{mcy}{"58}

\newcommand{\hinter}[2][]{\ensuremath{\widetilde{H}^{I,\infty}_{#1}\pqty{#2}}}

\newcommand{\peb}[0]{\ensuremath{\overline{p_{e}}}}
\newcommand{\esk}[0]{\ensuremath{\overline{\varepsilon}}}

\newcommand{\refcite}[2][]{{\IfSubStr{#2}{,}{Refs.}{Ref.}~\onlinecite[#1]{#2}}}

\let\orghat\hat
\def\hat#1{\orghat{\kern0pt #1}}

\crefname{section}{Sec.}{Secs.}

\begin{document}
\synctex=1
\title{Emulating non-Markovian system-bath dynamics with
  parametrically driven cavities}

\author{V. Boettcher \orcidlink{0000-0003-2361-7874}}
\email[]{valentin@physics.mcgill.ca}
\homepage[]{https://protagon.space}
\affiliation{Department of Physics, McGill University, Montréal, Québec, Canada H3A 2T8}
\author{F. Pellerin}
\email[]{felix.pellerin@umontreal.ca}
\affiliation{Département de Physique, Université de Montréal, C.P. 6128, Succursale Centre-Ville, Montréal, Québec H3C 3J7, Canada}
\author{P. St-Jean \orcidlink{0000-0002-2072-0756}}
\email[]{philippe.st-jean@umontreal.ca}
\affiliation{Département de Physique, Université de Montréal, C.P. 6128, Succursale Centre-Ville, Montréal, Québec H3C 3J7, Canada}
\author{W. A. Coish \orcidlink{0000-0001-8351-1014}}
\email[]{coish@physics.mcgill.ca}
\affiliation{Department of Physics, McGill University, Montréal, Québec, Canada H3A 2T8}

\date{\today}

\begin{abstract}
We introduce and characterize a scheme for emulating non-Markovian quantum system-bath dynamics using the discrete electromagnetic field modes of a parametrically driven cavity. In this scheme, the character of a bosonic bath (the form of the bath spectral density) is tied directly to the shape of the real-time waveform describing periodic parametric cavity modulation. As an explicit example, we demonstrate the feasibility of this scheme for a fiber-loop experiment with currently achievable parameters, supported by numerical simulations and analytical estimates. In particular, we show that a localization transition of the spin-boson model can be accurately emulated in this fiber-loop cavity system. This localization effect is characterized by a sharp transition from partial decay to complete decay for a two-level system coupled to a bosonic bath as the bath spectral density \(J(\omega)\propto \omega^{s}\) is continously tuned from the sub-ohmic (\(s<1\)) to the super-ohmic (\(s>1\)) regime. The transition is only exactly realized in the thermodynamic limit (for an infinite number of bath modes) and for sufficiently weak system-bath coupling. We highlight a competition between the weak-coupling and thermodynamic limits in this problem and show that challenges in approximately realizing the transition can nevertheless be overcome. Finally, we provide bounds on the systematic error introduced when emulating non-Markovian dynamics for arbitrary system observables. The scheme presented here can enable the realization of a modular platform for engineering custom non-Markovian baths, with potential applications in quantum thermodynamics, thermalization of many-body systems, and resource-efficient quantum simulations of open quantum systems.
\end{abstract}
\maketitle

\section{Introduction}
\label{sec:introduction}
Every system occurring in nature interacts with its environment~\cite{Breuer2002,Preskill2018}.  Often, the environment is modeled in an effective ``memoryless'' way via a Markov approximation.  The Markov approximation~\cite{Rivas2012,Breuer2002}, a prerequisite to the broadly applied~\cite{Blais2021,Harrington2019,Verstraete2009,Barreiro2011} Gorini--Kossakowski--Sudarshan--Lindblad formalism~\cite{Gorini1976,Lindblad1976}, is only valid under certain (sometimes highly restrictive) conditions~\cite{Fick1990,Rivas2012}.

There has been widespread interest in the effects of more general non-Markovian dynamics~\cite{Shrikant2023,Suarez1992,Laine2014,Heineken2021,Gulacsi2023,Ricottone2020,Yao2020,Boettcher2024,Wiedmann2020,Mukherjee2020,Yang2024a,Guo2021,khaetskii2002electron,Taylor2003,Anderson2004,Jobez2014,Terhal2005,Preskill2018,White2022,Mirkin2019,Laine2014}.  Non-Markovian descriptions include memory effects~\cite{Suarez1992,Laine2014,Heineken2021}, often leading to revivals~\cite{Gulacsi2023}, and these effects typically arise from a more complex energy structure of the environment beyond the broad-bandwidth and weak-coupling limits. Environments that lead to non-Markovian dynamics can either induce or destroy dynamical topological effects~\cite{Ricottone2020}. History-dependent (non-Markovian) dynamics also have implications for quantum thermodynamics~\cite{Boettcher2024,Wiedmann2020,Mukherjee2020}. Non-Markovian effects can enhance qubit readout \cite{Yang2024a} and can be exploited in a quantum memory~\cite{Guo2021,Taylor2003,Anderson2004,Jobez2014}. Finally, non-Markovian effects are relevant for fault tolerant quantum computing, potentially leading to detrimental hard-to-correct errors~\cite{Terhal2005,Preskill2018,White2022}.
On the other hand, the fact that non-Markovian processes have a ``memory'' can be exploited to reduce the impact of errors in special cases \cite{Kattemolle2023}.

Here, we introduce and characterize a simple scheme for emulating non-Markovian system-bath dynamics. This scheme can be realized either in the quantum regime or in classical systems that emulate quantum systems, such as parametrically driven cavities made from loops of optical fiber \cite{Dutt2019,Yang2022a,Senanian2023,Pellerin2024}. If experiments can be pushed to reach the criteria outlined in this paper, it may be possible to use the scheme presented here to experimentally learn about nontrivial non-Markovian dynamics in a regime that would otherwise be computationally difficult to access with currently available numerical approaches~\cite{Zhang2024,Gera2023,Dalton2001,Wang2010,Roden2012,Nusseler2020,Strathearn2018,Tanimura1990,Tang2015,Hartmann2021a,Suess2014,xu2026simulating}. In the shorter term, it may be possible to leverage non-Markovian effects that are known to occur in quantum systems, but to realize them in a classical analogue system with the goal of developing new technologies or functionality. This ``quantum mimetics'' is much weaker than the greater goal of realizing resource-efficient universal quantum simulation, but it can nevertheless provide inspiration for designing technologies with new functionality. This is similar in spirit to the field of biomimetics \cite{vincent2006biomimetics,bhushan2009introduction}, where phenomena seen in nature inspire new designs (e.g., the shape of a bird's wing inspires airplane design, or the texture of a gecko's feet inspires the development of new adhesives).

\begin{figure}[tb]
  \centering
  \includegraphics[page=2]{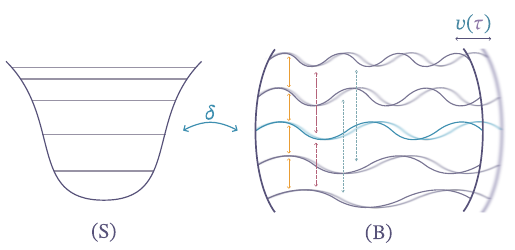}
  \caption{\label{fig:cartoon}A system (S)---here, an anharmonic oscillator---coupled with coupling strength \(\delta\) to a single mode of a parametrically modulated cavity that serves as a bath (\(\bath\)). Periodic parametric modulation with amplitude \(v(\tau)=v(\tau+T)\) couples the cavity modes.}
\end{figure}

The proposal presented here generalizes and extends a number of other recent works on parametrically driven classical or quantum harmonic modes that may emulate a nontrivial effective quantum dynamics \cite{ivakhnenko2018simulating,lorenz2023classical,bernazzani2024fluctuating,peyruchat2025landau,Vu2025}. More specifically, the scheme presented in this paper allows for a controlled classical or quantum apparatus to emulate the quantum dynamics induced by a bosonic bath with a highly tunable spectral density. This is achieved by parametrically driving a single electromagnetic cavity [labeled ``(\(\bath\))'' in \cref{fig:cartoon}]. Parametric driving of the cavity induces a coupling between its undriven eigenmodes. The dynamics of the driven cavity are then described by a one-dimensional tight-binding model with ``sites'' (the undriven eigenmodes) defined in a synthetic dimension and with long-range hopping amplitudes controlled by Fourier coefficients of the parametric drive amplitude~\cite{Dutt2019}. By tuning the hopping amplitudes, the band structure of the tight-binding model can be tailored to produce a desired density of states \(D(\varepsilon)\). In the simplest limit (illustrated in \cref{fig:cartoon}), a time-dependent periodic drive amplitude \(v(\tau)=v(\tau+2\pi/\Omega)\), with period determined by the cavity free spectral range \(\Omega/2\pi\), directly realizes a single band with energy dispersion \(\varepsilon(k)=v(\tau=k/\Omega)\), giving \(D(\varepsilon)\propto |d\varepsilon(k)/dk|^{-1}\). Tight-binding models with only short-range hopping produce a limited set of possibilities for the density of states (e.g. \(D(\varepsilon)\propto \varepsilon^{d/2-1}\) for a quadratic dispersion \(\varepsilon(k)\propto k^2\) in \(d \in \{1,2,3\}\) dimensions). In contrast, long-range hopping induced in the synthetic dimension \cite{Pellerin2024}  admits a near arbitrary dispersion, allowing for a low-energy density of states \(D(\varepsilon)\propto \varepsilon^{s}\), with \(s\) a continuous parameter. This density of states then determines a highly tunable spectral density \(J(\varepsilon)\propto D(\varepsilon)\) for the bath. When many cavity modes participate in the dynamics, even a single parametrically driven cavity can be used to accurately emulate a quasi-continuum of bath modes. This cavity can then be integrated as a ``bath module'' that can be coupled to a variety of systems (including qubits, anharmonic modes, or other cavities). The bath spectral density can be verified directly through a simple transmission experiment~\cite{Dutt2019} and (unlike the case of open quantum systems, where the bath is integrated out), here bath observables can be directly measured through Fourier analysis of the nonstationary cavity transmission~\cite{Pellerin2024}.

\begin{figure*}[tbp]
  \centering
  \includegraphics{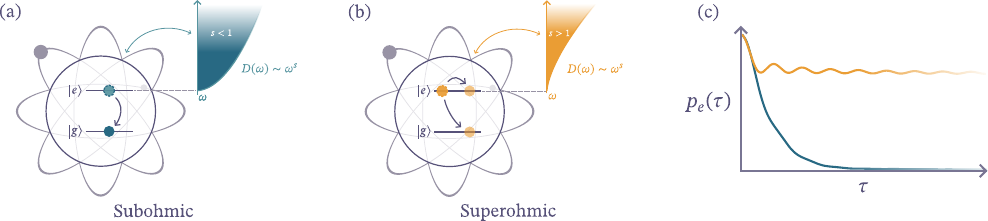}
  \caption{\label{fig:ww-cartoon} A two-level atom coupled to a bath having a spectral density $J(\omega)\propto \omega^s$ that is either  (a) subohmic ($s<1$), or (b) superohmic ($s>1$). (c) probability $p_e(\tau)$ for the atom to be found the excited state \(\ket{e}\) at time $\tau$. The atom completely decays into the ground state \(\ket{g}\) when coupled to a subohmic bath, whereas the atom only decays partially for a bath with a superohmic spectral density.}
\end{figure*}
The ideas described above can be applied directly to realize the well-known Wigner-Weisskopf model of atomic decay into a continuum \cite{Weisskopf1930,Scully1997}. As an important illustrative example, we therefore theoretically analyze the emulated Wigner-Weisskopf dynamics for a particular experimental setup based on classical light in parametrically modulated fiber-loop cavities~\cite{Dutt2019,Yang2022a,Senanian2023,Pellerin2024}. The Wigner-Weisskopf model provides a testbed for nontrivial non-Markovian effects as it exhibits a discontinuity in the long-time atomic decay probability, conditioned on the exponent \(s\) of the low-energy bath spectral density, \(J(\varepsilon)\propto \varepsilon^s\), in going from the ohmic or subohmic (\(s\le 1\)) [see \cref{fig:ww-cartoon} (a)] to the superohmic regime (\(s>1\)) [see \cref{fig:ww-cartoon}(b)]~\cite{Palma1997,Krinner2018,Leggett1987,Ricottone2020}.

Wigner-Weisskopf dynamics have previously been simulated in a cold-atom setup with a focus on a tuned transition from Markovian to non-Markovian dynamics induced by an energy offset to a continuum band \cite{Krinner2018}. In the present work, we theoretically explore Wigner-Weisskopf dynamics in a very different physical setting (driven cavities) and our emphasis is on a situation where the spectral density can be more finely tuned to describe a near arbitrary power law at low energy. In a different cold-atom experiment \cite{Lapp2019}, an effective bath has been created from a nearest-neighbor tight-binding chain with the goal of engineering Markovian loss rates. Our theoretical approach is conceptually similar in that we use a one-dimensional tight-binding chain, but in our approach this is achieved instead in a synthetic dimension defined by the undriven eigenmodes of a driven cavity, allowing for long-range hopping and a highly tunable bath spectral density. A recent theoretical proposal has also explored the idea of generating an effective bath in the synthetic dimension associated with a parametrically driven resonator \cite{Du2022}, but there the focus was on realizing the interesting physical effects enabled by a giant atom rather than tuning the spectral density of the bath. A recent experiment based on the manipulation of many light modes undergoing a transformation controlled by spatial light modulators and a nonlinear crystal has simulated the dynamics of a system interacting with an effective bath, described by a network \cite{Renault2023}. The experiment of \refcite{Renault2023}, based on an earlier theoretical proposal \cite{Nokkala2018}, focuses on simulating a bath with a spectral density that is determined by a particular network giving the bath Hamiltonian. Our work, in contrast, focuses on the problem of designing a bath with a given predefined and tunable spectral density. To this end, we introduce a simple procedure to design the characteristic low-frequency spectral density of the bath in a different platform (driven linear cavities).

Trapped ions are an attractive platform for simulating system-bath models due to their highly controllable discrete internal (e.g. spin or orbital) and motional (bosonic) degrees of freedom.
There has been good progress in developing both proposals \cite{Porras2008,Lemmer2018} and experiments \cite{Wang2024c,Than2025,Sun2025} in this area. However, each of these proposals or experiments has so far achieved a limited tunability of the spectral density. For example, in \refcite{Porras2008} the authors proposed to realize a spin-boson model where the bath arises from the phonon degrees of freedom of a long chain of (\(\sim 50\)) ions. For the one-dimensional implementation in  \refcite{Porras2008}, the spectral density was limited to \(J(\omega)\propto \omega^{s}\) where \(s=\pm 1\). A similar idea was used in the experiment presented in \refcite{Wang2024c}, which realized a spin-boson model with a chain of up to 20 trapped ions. The observed spectral density in this experiment was shown to be highly variable, but without fine control over the spectral exponent $s$ determining the low-frequency behavior. More refined control over the spectral density has been demonstrated in \refcite{Sun2025} (based on the theoretical work in \refcite{Lemmer2018}, where controlled bosonic dephasing provides a greater level of tunability). The authors of \refcite{Sun2025} demonstrated the versatility of their method by simulating a spectral density \(J(\omega)\propto \omega^{s}\) with spectral exponents \(s=0.5,1.0,2.0\). However, the frequency range over which the spectral density could be reproduced was limited by the fact that only three motional modes of the trapped ions were involved.

The goal in this paper is to emulate a spin-boson model with a tunable bath spectral density \(J(\omega)\propto \omega^{s}\), where, in contrast to previous works, the parameter $s$ can be continuously tuned to almost any value. The quality of the spectral density that can be realized here depends predominantly on the number accessible cavity modes $N$ and on the bandwidth of a parametric cavity drive, relative to the cavity free spectral range (determining the number of discrete Fourier coefficients $N_F$ in the drive). The number $N$ can be very large in certain implementations (e.g., $N>10^4$ for optical fiber loop cavities~\cite{McMahon2023}). The number of drive Fourier coefficients $N_F$ can also be very large ($N_F\gg 1$) if the drive is generated by a high-quality waveform generator. The large achievable values of $N$ and $N_F$ can lead to significant advantages over strategies that rely on a small number of less controllable bosonic degrees of freedom.

The rest of this paper is organized as follows. In \Cref{sec:open-quantum-systems}, we introduce some basic terminology and review system-bath models. In \Cref{sec:from-long-range} we outline the strategy used to emulate a bath with an arbitrary spectral density from a parametrically modulated cavity. In \Cref{sec:appl-cond-wign}, we consider an explicit experimental setup involving parametrically modulated optical fiber loops and we show that this setup can emulate a transition from partial decay to complete decay of an atom coupled to a bosonic bath (the Wigner-Weisskopf model) as the bath spectral density is tuned. Finally, in \Cref{sec:fidelity-finite-size}, we discuss the accuracy of this approach in approximately emulating dynamics for an infinite (continuum) bath with a finite bandwidth of the modulation and a finite number of available cavity modes. We conclude in \cref{sec:conclusion}.

\section{System-Bath Model}
\label{sec:open-quantum-systems}
A typical system-bath model~\cite{Leggett1987,Caldeira2014,Hartmann2021a} is given by the following Hamiltonian:
\begin{gather}
  \label{eq:system-bath-generic}
  H =H_{\sys}+ H_{\bath}+ H_{\inter}, \\
  \begin{aligned}[t]
    \label{eq:bath-part}
    H_{\bath} & =\sum_{n=1}^{N}\omega_{n}b_{n}^{\dag}b_{n}, & H_{\inter} & = \delta\bqty{L^\dag B + L B^\dag},
  \end{aligned}\\
  \begin{aligned}[t]
    \label{eq:bathop}
    B & \equiv \sum_{n=1}^{N}{g}_{n}b_{n}, & \comm{b_{n}}{b_{m}^\dag} & =\delta_{nm},
  \end{aligned}
\end{gather}
where \(b_n\) is an annihilation operator for bosonic mode \(n\) of the bath, having frequency \(\omega_n \ge 0\), the system Hamiltonian \(H_{\sys}\) is arbitrary, and \(L\) is a system operator that couples to the collective bath operator \(B\) with coupling strength \(\delta\). The coupling constants \(g_n\) are assumed to be dimensionless and normalized such that \(\sum\abs{g_{n}}^{2}=1\).

It is advantageous to remove the bath Hamiltonian \(H_{\bath}\) through a unitary transformation (setting \(\hbar=1\)),
\begin{equation}
  \label{eq:system-bath-generic-interaction}
  \begin{gathered}
    H(\tau) \equiv \eu^{\iu H_{\bath}\tau}H\eu^{-\iu H_{\bath}\tau } - H_{\bath} =  H_{S} + H_{\inter}(\tau),\\
  \end{gathered}
\end{equation}
where
\begin{gather}
  H_{\inter}(\tau)  \equiv  \delta\bqty{L^\dag B(\tau) + L B^\dag(\tau)}, \\
  B(\tau)  \equiv \sum_{n=1}^{N}{g}_{n}b_{n}\eu^{-\iu \omega_{n}\tau}\label{eq:b-stationary}.
\end{gather}
For a bath initially at zero temperature, the effect of the bath on the system is determined solely by the bath correlation function \cite{Rivas2012,Roden2012,Huang2024}
\begin{equation}
  \label{eq:bath-time-correlation}
  c(\tau) = \ev{B(\tau)B^\dag(0)}{0} = \sum_{n}\abs{g_{n}}^{2}\eu^{-\iu \omega_{n}},
\end{equation}
with the vacuum state of the bath \(\ket{0}\). In the convention chosen here, \(c(\tau)\) is dimensionless and \(c(0)=1\). All information about the bath dynamics is equivalently contained within the bath spectral density
\begin{equation}
  \label{eq:spectral-density-decomposition}
  J(\omega)\equiv \frac{1}{2}\int_{-\infty}^\infty\dd{\tau} c(\tau)\eu^{\iu \omega \tau}= {\pi} \sum_{n}\abs{{g}_{n}}^{2}\delta(\omega-\omega_{n}),
\end{equation}
which can be replaced by a smooth function of frequency in the continuum limit. After taking the continuum limit, the dynamics may become effectively irreversible, whereas for a finite number of bath modes \(N\), Poincaré recurrences are always present~\cite{Weiss2012}.

A designed spectral density \(J(\omega)\) [or, equivalently, the correlation function \(c(\tau)\)] can be realized by manipulating either the coupling parameters \(g_n\), the frequencies \(\omega_n\), or both. In the analysis that follows, we focus on a strategy where the \(g_n\) are constant and the density of states can be tuned to realize a particular quasi-continuum distribution of \(\omega_n\). This situation is naturally realized for the physical setup that we consider in the following section.

\section{Emulating System-Bath Models with a Driven Cavity}
\label{sec:from-long-range}
\begin{figure*}[t]
  \centering
  \includegraphics[width=\textwidth,page=1]{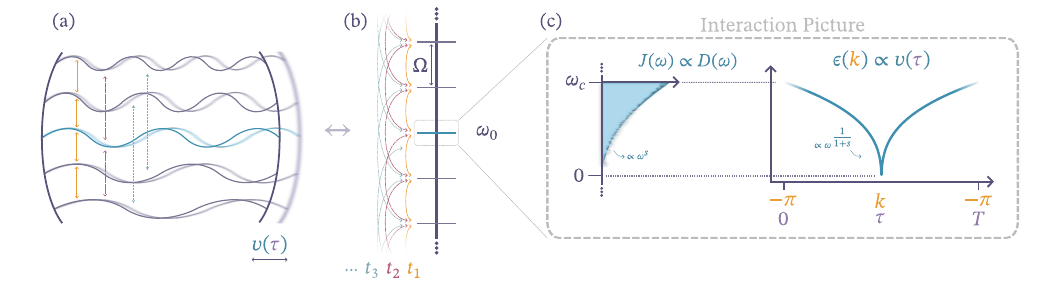}
  \caption{\label{fig:totalscheme} (a) A cavity modulated by a
    parametric drive \(v(\tau)\) featuring evenly spaced frequency modes
    centered around \(\omega_{0}\) with spacing \(\Omega\). (b) The drive
    realizes a hopping model with long-range hopping amplitudes
    \(t_{l}\). (c) In the interaction picture, the bare modes become
    a quasi-continuum with a spectral density \(J(\omega)\) derived from a density of states \(D(\omega)\), here chosen to have a low-frequency
    power-law form \(\propto \omega^{s}\) with spectral exponent \(s\) and
    cutoff frequency \(\omega_{c}\). The
    density of states arises from the dispersion relation \(\varepsilon(k)\)
    [\(0\leq \varepsilon(k)\leq \omega_{c}\), (\(-\pi< k\leq \pi\))] for the long-range hopping model through \cref{eq:inverse-density-relationship}. The dispersion relation is
    itself directly related to the
    drive amplitude \(v(\tau)\) with period \(T\) [\cref{eq:epsilon-approximate-dispersion}].}
\end{figure*}

In this section, we consider a method for emulating the system-bath model described in \cref{sec:open-quantum-systems} with a driven cavity  (the bath, \cref{fig:totalscheme}) coupled to a nonlinear oscillator (the system). The basic idea behind this scheme is to realize an effective one-dimensional long-range hopping model in the driven cavity. The dispersion relation, and thus the density of states, can be tuned by adjusting the long-range hopping amplitudes. This fact can be exploited to realize a tunable spectral density \(J(\omega)\).

\begin{figure}[t]
  \centering
  \includegraphics{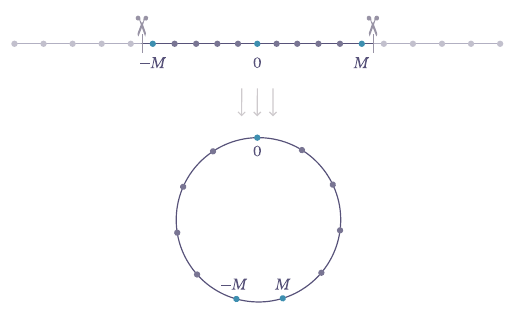}
  \caption{\label{fig:boundary-sketch} Top: schematic illustration of `sites' (in a synthetic dimension) describing cavity modes. We truncate the spectrum of cavity modes to a finite number (\(N=2M+1\)). After truncating the set of modes, we assume periodic boundary conditions for convenience.}
\end{figure}

A generic time-dependent Hamiltonian describing a parametrically driven cavity coupled to a nonlinear oscillator in a concrete experimental realization is given by
\begin{equation}
  \label{eq:system-hamiltonian-components}
  \widetilde{H}_{\experimental}(\tau)= {\widetilde{H}_{0}} + \widetilde{H}_{\sys}  +  \widetilde{H}_{\bath}(\tau) + \widetilde{H}_{\inter}.
\end{equation}
We use a tilde ($\, \widetilde{\,} \,$) to distinguish Hamiltonians for some physical implementation from the idealized versions described above. The Hamiltonian \({\widetilde{H}}_{0}\) will be used to define the interaction picture, the system Hamiltonian describing a nonlinear oscillator is \(\widetilde{H}_{\sys}\), the Hamiltonian of the driven (bath) cavity is \(\widetilde{H}_{\bath}(\tau)\), and the interaction between the driven cavity and the nonlinear oscillator is \(\widetilde{H}_{\inter}\):
\begin{align}
  \label{eq:systems-decomposed-hamiltonian}
  {\widetilde{H}_{0}}         & \equiv \omega_{0}a^{\dag}a  + {\sum_{n=-M}^{M} \pqty{\omega_{0}+n \Omega} {b}_{n}^{\dag}{b}_{n}},                              \\
  \widetilde{H}_{\sys}        & \equiv \omega_{a}a^{\dag}a - \frac{\omega^{(2)}_{a}}{2}a^\dag a^\dag aa,                         \label{eq:system-hamiltonian} \\
  \widetilde{H}_{\bath}(\tau) & \equiv \sum_{n,m}v_{n-m}(\tau){b}_{n}^{\dag}{b}_{m},                          \label{eq:modulation-hamiltonian}                \\
  \widetilde{H}_{\inter}      & \equiv {\delta} \pqty{a^{\dag}{b}_{0}+\hc}\label{eq:inter-ham},
\end{align}
where \(\comm{b_{m}}{b_{n}^\dag} = \delta_{mn}\), \(\comm{a}{b_{n}} = 0\), and \(\comm{a}{b_{n}^{\dag}} = 0\). Here, the free spectral range \(\Omega/2\pi\) sets the spacing between cavity modes. We assume that only a finite number of cavity modes may be excited with appreciable amplitude, allowing us to truncate the cavity to \(N=2M+1\) modes $n=-M \ldots M$, over which $\Omega$ is approximately constant. These modes represent the sites in a synthetic dimension for an effective hopping model~\cite{Yuan2016,Yuan2018,Dutt2019}. The frequency of the central mode ($n=0$) is $\omega_0$ and the system oscillator is detuned from $\omega_0$ by \(\omega_{a}\). The parameter \(\omega^{(2)}_{a}\) controls the anharmonicity of the system, and the parameters \(v_{n-m}(\tau)\) are drive amplitudes. The form of the drive amplitudes \(v_{nm}(\tau)=v_{n-m}(\tau)\) assumes that the drive acts equivalently on every pair of cavity modes that are separated by the same frequency.
We also assume these amplitudes to be periodic in time,
\begin{equation}
  \label{eq:drive-period}
  v_{n-m}(\tau) = v_{n-m}(\tau+T);\quad T= \frac{2\pi}{\Omega},
\end{equation}
and we set the mean value of the drive amplitudes to zero for \(m=n\),
\begin{equation}
  \label{eq:drive_mean_zero}
  \int_{0}^{T}\dd{\tau}v_{0}(\tau) = 0,
\end{equation}
as any finite value could simply be absorbed into the definition of the center frequency
\(\omega_{0}\).
Since, by assumption, much fewer than \(M\) modes will be excited with appreciable amplitude, we ignore boundary effects and, for convenience, we choose periodic boundary conditions as illustrated in \cref{fig:boundary-sketch},
\begin{equation}
  \label{eq:periodic-boundaries}
  {b}_{M+1}={b}_{-M}.
\end{equation}

We now adopt an interaction picture with respect to the Hamiltonian \(\widetilde{H}_{0}\). In addition, we retain the restriction $n\in [-M,M]$ on the index $n$ in \cref{eq:modulation-hamiltonian}, but we allow the difference $l=n-m$ to be formally unrestricted. This approximation is valid provided \(v_{l}\) decays to a negligible value when \(l>l_{0}\) for some \(l_{0} \ll M\) and if the modulation \(v_{l}(\tau)\) only contains Fourier components with frequencies much smaller than \(M\Omega\). This procedure leaves us with:
\begin{gather}
  \label{eq:interaction-picture-transformation}
  \begin{aligned}
    \hinter{\tau} & \equiv \widetilde{H}_{\bath}^{I,\infty}(\tau) +\widetilde{H}_{\inter} + \widetilde{H}_{\sys},
  \end{aligned}\\
  \label{eq:bath_rotframe}%
  \hinter[\bath]{\tau} \equiv {\sum_{n=-M}^{M}\sum_{l=-\infty}^{l=\infty} v_{l}(\tau) \eu^{\iu \Omega l \tau} b_{n+l}^\dag b_{n}}.
\end{gather}
Here, we have introduced the superscript "\(^\infty\)" to indicate that the index $l$ has been extended to an infinite range. Conveniently, \(\hinter[\bath]{\tau}\) has translational invariance with respect to the mode index \(n\).
Thus, it will be diagonal when expressed in terms of Bloch modes \cite{Dutt2019}
\begin{equation}
  \label{eq:mode-fourier-transformation}
  \begin{multlined}
    b_{k}=\frac{1}{\sqrt{N}} \sum_{n=-M}^{M} b_{n}\eu^{-\iu k n}, \\
    k = m\frac{2\pi}{N},\, m=-M, -M + 1,\ldots,M.
  \end{multlined}
\end{equation}
In terms of these modes, we can rewrite the approximate bath Hamiltonian as
\begin{align}
  \label{eq:bath-translation-invariance}
  \hinter[\bath]{\tau}            & = \sum_{k} \widetilde{\varepsilon}_k(\tau)  b_{k}^\dag b_{k},                                \\
  \widetilde{\varepsilon}_k(\tau) & = \sum_{l} \eu^{-\iu l (k - \Omega \tau )} v_{l}(\tau)\label{eq:time-depdendent-dispersion}.
\end{align}

We now remove the free evolution of the bath through the transformation [cf.~\cref{eq:system-bath-generic-interaction}]:
\begin{equation}
  \label{eq:final-cavity-bath}
  \begin{aligned}
    \widetilde{H}(\tau) & \equiv \eu^{\iu\int_{0}^{\tau}\dd{\tau'} \hinter[\bath]{\tau'}}\hinter{\tau}\eu^{-\iu\int_{0}^{\tau}\dd{\tau'} \hinter[\bath]{\tau'}}-\hinter[\bath]{\tau}, \\
                        & = \widetilde{H}_{\sys} + \widetilde{H}_{\inter}(\tau),
  \end{aligned}
\end{equation}
where
\begin{align}
  \widetilde{H}_{\sys}         & =\omega_{a}a^{\dag}a - \frac{\omega^{(2)}_{a}}{2}a^\dag a^\dag aa,                                               \\
  \widetilde{H}_{\inter}(\tau) & = \delta \bqty{a^{\dagger} \widetilde{B}(\tau) + \hc},\label{eq:H_I}                                             \\
  \widetilde{B}(\tau)          & =\frac{1}{\sqrt{N}} \sum_{k}b_{k}\eu^{-\iu \bqty{\varepsilon_{k}\tau +\phi_{k}(\tau)}}.\label{eq:Bnonstationary}
\end{align}
If we make the association \(g_n\to g_k\) for the coupling constants in \cref{eq:b-stationary}, then here \(g_{k}=1/\sqrt{N}\), independent of the index \(k\).  In \cref{eq:Bnonstationary}, above, we have introduced the time-averaged dispersion relation
\begin{equation}
  \label{eq:average-dispersion-calculation}
  \varepsilon_{k}  \equiv\frac{1}{T} \int_{0}^{T}\dd{\tau}\widetilde{\varepsilon}_k(\tau),
\end{equation}
and the nonstationary (explicitly $\tau$-dependent) phase
\begin{equation}
  \label{eq:residual-phase}
  \phi_{k}(\tau)\equiv \int_{0}^{\tau }\dd{\tau'}\bqty{\widetilde{\varepsilon}_k(\tau')-\varepsilon_{k}}.
\end{equation}
For a quantum system that is well described by an anharmonic oscillator, or where the system can be approximately described in some low-dimensional subspace of an anharmonic oscillator, we can make the direct identification between the abstract model system Hamiltonian $H_\sys$ and the experimental realization $\widetilde{H}_\sys$: $H_\sys\simeq \widetilde{H}_\sys$.

The system-bath Hamiltonian in \cref{eq:final-cavity-bath} has a similar form to the target system-bath model [\cref{eq:system-bath-generic-interaction}], but here, the bath operators \(B(\tau)\) evolve generally under a nonstationary process. In this case, the system dynamics are more generally determined by a two-time bath correlation function
\begin{equation}
  \label{eq:bcf-nonstat}
  \begin{aligned}
    c(\tau,\tau') & = \ev{\widetilde{B}(\tau)\widetilde{B}^{\dagger}(\tau')}                                            \\
                  & = \frac{1}{N}\sum_{k}\eu^{-\iu \bqty{{\varepsilon}_{k}\pqty{\tau -\tau'}+ {\phi}_{k}(\tau)-\phi_{k}(\tau')}}.
  \end{aligned}
\end{equation}
Although the nonstationary phase vanishes stroboscopically at integer multiples of the period, \(\tau=nT\), deviations from \(\phi(\tau)=0\) at intermediate times can generally accumulate in observables constructed from the system or bath operators. To accurately emulate a system-bath evolution for a stationary (time-local) bath correlation function, it is advantageous to work in a regime where the nonstationary phase is a small correction. Under the condition \(\abs{\varepsilon_{k}} \ll \Omega\) (the usual rotating-wave approximation), we find \(\abs{\phi_{k}(\tau) - \phi_{k}( \tau')} \ll 1\) for all \(\tau,\tau'\) (see also the discussion in \cref{sec:rotat-wave-appr} below for a more detailed argument).  In the following, we assume this condition holds and neglect the nonstationary phase \(\phi_{k}\), allowing for an emulation of the time-local bath correlation function
\begin{equation}
  \label{eq:bcf-stat}
  c(\tau)=\frac{1}{N}\sum_{k}\eu^{-\iu \varepsilon_{k} \tau}.
\end{equation}
The character of this correlation function and the associated spectral density \(J(\omega)=(\pi/N)\sum_n\delta(\omega-\varepsilon_k)\) can be varied by tuning the dispersion \(\varepsilon_{k}\).

In \cref{sec:deta-disc-conv}, corrections are given to the correlation function (and hence to the spectral density) that go beyond the rotating-wave approximation [due to the nonstationary phase $\phi_k(\tau)$] and that account for a more general distribution of coupling parameters $g_n$.

\subsection{Tuning the dispersion relation}
\label{sec:tun-disp-relat}
Due to the assumed periodicity of the drive amplitudes [\cref{eq:drive-period}], they can be Fourier expanded to
\begin{equation}
  \label{eq:transformed-fourier-drive}
  v_{l}(\tau)=\sum_{j=-N_{F}}^{N_{F}}v_{l}^{j}\eu^{-\iu j \Omega \tau},
\end{equation}
where \(N_{F}\) controls the bandwidth \(\sim N_{F}\Omega\) of the drive.

Neglecting the nonstationary phase \(\phi_{k}(\tau)\), as discussed above, is equivalent to time-averaging the interaction-picture Hamiltonian in \cref{eq:interaction-picture-transformation} over one period. The effective tight-binding model realized in the driven cavity thus becomes apparent:
\begin{equation}
  \label{eq:cav-h-averaged}
  \frac{1}{T}\int_{0}^{T}\dd{\tau}\hinter[\bath]{\tau}
  = {\sum_{n=-M}^{M}\sum_{l=-N_{F}}^{l=N_{F}} t_{l}b_{n+l}^\dag b_{n}}
  = \sum_{k} \varepsilon_k  b_{k}^\dag b_{k}.
\end{equation}
Above, we have used \cref{eq:transformed-fourier-drive} and introduced the hopping amplitudes
\begin{equation}
  \label{eq:hopping-amps-def}
  t_{l} \equiv v^{l}_{l},
\end{equation}
to obtain the time-averaged dispersion relation [\cref{eq:average-dispersion-calculation}],
\begin{equation}
  \label{eq:fourier-dispersion}
  \varepsilon_{k} = \sum_{l=-N_{F}}^{N_{F}} t_{l}\eu^{-\iu l k}.
\end{equation}
The dispersion \(\varepsilon_{k}\) is assumed to have a finite bandwidth \(\omega_c\),
\begin{equation}
  \label{eq:cutoff-def}
  0\leq \varepsilon_{k} \leq \omega_{c} \leq \sum_{l}\abs{t_{l}}.
\end{equation}
When the eigenvalues \(\varepsilon_k\) are treated as a bath spectrum, \(\omega_c\) sets the cutoff frequency.

In the special case where the drive amplitude is independent of the frequency difference between cavity modes, \(v_{l}(\tau)=v(\tau)\), it follows from \cref{eq:transformed-fourier-drive,eq:fourier-dispersion} that the dispersion is directly given by the drive amplitude
\begin{equation}
  \label{eq:epsilon-approximate-dispersion}
  \varepsilon_{k}=\varepsilon(k) = v(\tau={k}/{\Omega}).
\end{equation}
This is the situation illustrated in \cref{fig:totalscheme}(c). In the thermodynamic limit (\(N\to \infty\)) the dispersion relation becomes a function of a continuous variable \(k\) [\(\varepsilon_k\to\varepsilon(k)\)], which is related directly to the continuous function of time $v(\tau)$ through \cref{eq:epsilon-approximate-dispersion}.

\subsection{Tuning the spectral density}
For the local (single-mode) coupling between system and bath assumed here, the collective bath operator \(B\) is a uniform linear combination of plane-wave mode operators \(b_{k}\) [\cref{eq:Bnonstationary}], equivalent to taking a system-bath model with uniform system-bath coupling, \(g_k=1/\sqrt{N}\). The spectral density is then directly proportional to the density of states \(D(\omega)\):
\begin{equation}
  \label{eq:josephson-density-function}
  J(\omega) = \frac{\pi}{N} \sum_{k} \delta(\omega-\varepsilon_{k}) \equiv {\pi} D(\omega).
\end{equation}
In the thermodynamic (continuum) limit, the spectral density can be obtained from the derivative of the dispersion
\begin{equation}
  \label{eq:inverse-density-relationship}
  J(\omega)=\pi D(\omega) = \frac{1}{2}\sum_{{k;\;{\varepsilon({k})=\omega}}}{\abs{{\dv{\varepsilon({k})}{k}}}}^{-1}.
\end{equation}
This allows, in principle, for an arbitrary spectral density \(J(\omega)\) given the appropriate dispersion relation \(\varepsilon(k)\).

\begin{figure}[t]
  \begin{multifig}
    \begin{subfigure}{\columnwidth}{\label{subfig:idealdisp}}
      \includegraphics{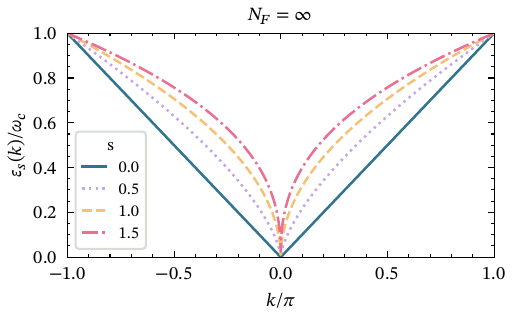}
    \end{subfigure}
    \begin{subfigure}{\columnwidth}{\label{subfig:lowpassdisp}}
      \includegraphics{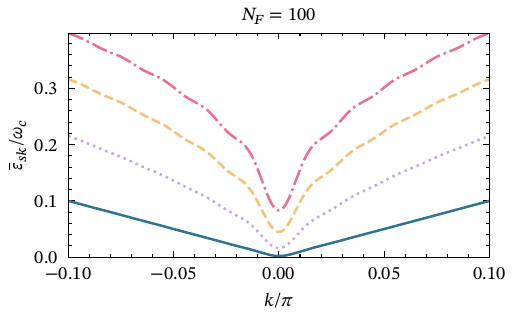}
    \end{subfigure}
  \end{multifig}
  \caption{\label{fig:dispersion-plot}Bath dispersion for various spectral exponents \(s\). \subref{subfig:idealdisp} The ideal dispersion relation, \cref{eq:dispersion-powerlaw}, which can be realized in the limit of an infinite number of Fourier coefficients, $N_F\to\infty$. \subref{subfig:lowpassdisp} Filtered dispersion [\cref{eq:filtered_bath_energies}] for $N_F=100$, shown over a smaller range of $k$ around $k\simeq 0$ to highlight details.}
\end{figure}
As a concrete example, we focus on a power-law spectral density with cutoff frequency \(\omega_{c}\ll \Omega\) and spectral exponent \(s > 0\):
\begin{equation}
  \label{eq:power-law-spectral-density}
  J(\omega) = \pi D(\omega) =
  \begin{cases}
    \pi \frac{(1+s)}{\omega_{c}^{s+1}} \omega^{s} & 0\leq \omega\leq \omega_{c} \\
    0                                             & \text{otherwise}.
  \end{cases}
\end{equation}
The dispersion relation yielding \cref{eq:power-law-spectral-density} is
given by
\begin{equation}
  \label{eq:dispersion-powerlaw}
  \varepsilon(k)=\varepsilon_s({k}) = \omega_{c}\pqty{\frac{\abs{k}}{\pi}}^{\frac{1}{s+1}}
  \qq{for} -\pi\leq k\leq \pi.
\end{equation}
The dispersion $\varepsilon_s({k})$ is illustrated in \cref{fig:dispersion-plot}(a) for various $s$. This dispersion can be realized straightforwardly through a time-dependent periodic modulation of the drive amplitude $v(\tau)$ [\cref{eq:epsilon-approximate-dispersion}], but for a general exponent $s$, a large number of Fourier coefficients $N_F$ may be necessary to accurately represent the sharp features near $k=0$ [see \cref{fig:dispersion-plot}(b)]. The specific condition for a sufficiently large $N_F$ depends on the observable of interest, but we find conditions to see nontrivial dynamics that are achievable in experiment, as discussed in \cref{sec:effects-finite-drive,sec:fidelity-finite-size}, below.

The effective hopping amplitudes $t_l$ [and hence the drive amplitudes, \cref{eq:hopping-amps-def}] that would reproduce the required bath energies for $N_F\to\infty$ can be obtained through:
\begin{equation}
  \label{eq:hop-amp}
  t_{sl} = \frac{1}{2\pi} \int_{-\pi}^{\pi}\dd{k}\varepsilon_s(k)\eu^{\iu k l}.
\end{equation}
Filtered bath energies that approximately reproduce the required dispersion for a finite number of nonzero Fourier coefficients $N_F$ can then be found from the amplitudes given in \cref{eq:hop-amp} through \cref{eq:fourier-dispersion}:
\begin{equation}\label{eq:filtered_bath_energies}
  \esk_{sk} = \sum_{l=-N_F}^{N_F} t_{sl} e^{-ikl}.
\end{equation}
The approximate (filtered) bath energies $\esk_{sk}$ with $N_F=100$ Fourier coefficients are shown in \cref{fig:dispersion-plot}(b). Explicit expressions for the hopping amplitudes resulting from the  dispersion relation in \cref{eq:dispersion-powerlaw} can be found in \cref{sec:dispersion}, \cref{eq:hypergeometric-coefficients-formula}.

 Notably, a similar power-law dispersion arising from a long-range hopping model was also invoked in \cite{Jiang2016} as a way to tune the low-energy single-particle dispersion of a resonant Bose gas to generate long-lived states, suppressing few-body loss mechanisms.

\subsection{Probing the driven cavity}
\label{sec:meas-spectr-dens}
\begin{figure}[t]
  \centering
  \includegraphics{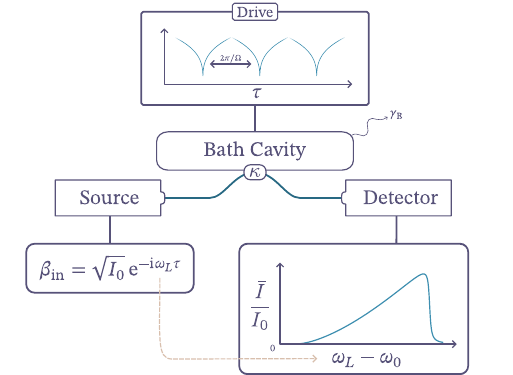}
  \caption{\label{fig:io_schematic} Schematic setup for verification of the spectral density. A coherent source (e.g.~laser) with frequency \(\omega_{L}\) feeds into a transmission line that is weakly coupled with strength $\kappa$ to the bath cavity mode having frequency \(\omega_{0}\). The bath modes are assumed to decay with a total photon loss rate \(\gamma=\kappa+\gamma_{\bath}\) through the transmission line (rate $\kappa$) and through intrinsic losses (rate $\gamma_B$). The output intensity results from an interference of the source field and the field exiting the bath cavity, and so its time-average $\bar{I}$ is related to the density of states realized in the bath [\cref{eq:intensity-response-compared}].}
\end{figure}
By probing the bath cavity with a coherent signal and measuring its transmission or reflection, one can probe the density of states~\cite{Dutt2019,Pellerin2024} and thus determine the spectral density to verify any experimental implementation and calibrate the hopping amplitudes \(t_{l}\).

For simplicity, we assume the bath cavity is decoupled from the system (\(\delta = 0\)). We then consider a setup (illustrated in \cref{fig:io_schematic}) in which the bath cavity (with loss rate \(\gamma_\bath\)) is coupled to a transmission line with coupling strength \(\kappa \ll \Omega\). This transmission line is fed by a coherent source such as a laser and terminates in a high-bandwidth detector that measures the time-resolved intensity of the electromagnetic field propagating along the transmission line.

We assume the coherent input field has an intensity \(I_{0}\). The output intensity \(I(\tau)\) after transients have decayed (\(\tau\gg1/\gamma_{\bath}\)) is then given by \cite{Dutt2019,Pellerin2024}
\begin{equation}
  \label{eq:signal-resonant-condition}
  \begin{aligned}
    \frac{I(\tau)}{I_{0}} & \approx 1 - \frac{\kappa}{2}\frac{\gamma}{\bqty{\varepsilon({\Omega \tau})-\Delta\omega}^{2} +\frac{\gamma^{2}}{4}},
  \end{aligned}
\end{equation}
where \(\gamma=\gamma_{\bath}+\kappa\) is the total loss rate of the bath cavity and \(\Delta\omega=\omega_{L}-\omega_{0}\) is the detuning of the frequency of the input field $\omega_{L}$ (due, e.g., to a laser) from the central frequency $\omega_{0}$.
As demonstrated experimentally in \refcite{Pellerin2024,Dutt2019}, the transmission is peaked around \(\varepsilon(\Omega \tau) = \Delta\omega\), allowing the dispersion relation \(\varepsilon(k)\) to be mapped out.

Averaging the output intensity, \cref{eq:signal-resonant-condition}, over one period, \(T=2\pi / \Omega\), gives
\begin{equation}
  \label{eq:intensity-response-compared}
  \frac{\bar{I}}{I_{0}} = \frac{1}{T}\int_{0}^{T}\dd{\tau} \frac{{I(\tau)}}{I_{0}}                                                                       = 1 - \frac{{\gamma\kappa}}{2}  \int_{0}^{\omega_{c}}\dd{\varepsilon}\frac{D(\varepsilon)}{\bqty{\varepsilon-\Delta\omega}^{2} +\frac{\gamma^{2}}{4}},
\end{equation}
where \(D(\varepsilon)\) is the density of states as defined in \cref{eq:josephson-density-function}. When the density of states is slowly varying  on the scale \(\gamma\), the Lorentzian can be approximately replaced by a delta function, giving \(\bar{I}/I_0\simeq 1-\pi\kappa D(\Delta\omega)\). The time-averaged intensity as a function of \(\Delta\omega\) then provides a direct measurement of the density of states and thus the spectral density.

\section{Experimental proposal to emulate Wigner-Weisskopf dynamics}
\label{sec:appl-cond-wign}
An important model that captures non-Markovian features such as revivals and persistent oscillations~\cite{Cohen-Tannoudji1977,Palma1997,Grabert1987,Krinner2018,Leggett1987,Ricottone2020} is the Wigner-Weisskopf model. The model, which is related to the Caldeira-Leggett model for a two-level system coupled to a bath \cite{Leggett1987}, describes a two-level atom decaying into a continuum of bath modes~\cite{Scully1997}.

In this section, we demonstrate how the Wigner-Weisskopf model can be emulated using classical light in parametrically modulated fiber-loop cavities. This can be done by exploiting the relationship between the coherent-state amplitudes associated with the parametrically driven linearly coupled two-cavity system discussed above and the quantum probability amplitudes associated with a two-level system interacting with a bath. We begin by proposing an experimental setup that can be used to emulate the Wigner-Weisskopf model in \cref{sec:experimental-setup}. Subsequently, in \cref{sec:mapping-to-ww} we discuss how to map this setup to the Wigner-Weisskopf model and we review a ``localization'' transition analogous to that seen in the Caldeira-Leggett model~\cite{Grabert1987}: For a low-energy spectral density \(J(\omega)\propto \omega^s\),  [cf. \cref{eq:power-law-spectral-density}], the atom decays fully when the spectral density is ohmic or subohmic ($s\le 1$) and it decays only partially when the spectral density is superohmic ($s>1$)~\cite{Palma1997,Ricottone2020}.
Finally, in \cref{sec:constraints} we discuss how this behavior of the model can be reproduced in an experiment, accounting for physical constraints of the proposed setup (including a finite number of cavity modes \(\sim N\), a finite number of nonzero Fourier coefficients \(\propto N_\mathrm{F}\), as well as finite loss rates).

\subsection{Proposed experimental setup}
\label{sec:experimental-setup}

As a realization of the bath cavity, we consider a ring cavity created by forming a loop of circumference \(L\) out of optical fiber. The eigenmodes of the loop then have the following wavenumbers and angular frequencies
\begin{equation}
  \label{eq:fiber-oscillator-frequencies}
  \begin{aligned}
    k_{m} & = (m_{0} + m)\frac{2\pi}{L}, & \omega_{m} & = \frac{c}{n_0}\abs{k_m},
  \end{aligned}
\end{equation}
where \(c\) is the speed of light in vacuum, \(n_{0}\approx 1.5\) is the refractive index of the fiber, and \(m_{0},m\in \ZZ\). The central frequency \(\omega_0\) is identified by the mode index \(m_{0} = \omega_{0} / \Omega\), and the mode spacing (free spectral range) is
\begin{equation}
  \label{eq:fsrdef}
  \frac{\Omega}{2\pi}\equiv \frac{c}{n_{0} L}.
\end{equation}

Parametric modulation of the cavity can be achieved with an electro-optic modulator (EOM) of length \(d\ll L\) inserted into the fiber loop. Such a modulator varies the optical length of the cavity. When \(\omega_{0}\gg M\Omega\), the amplitude of the modulation coupling modes $m, m'$, $v_{m,m'}(\tau)\simeq v_{l}(\tau)$, becomes a function only of the mode frequency difference $l=m-m'$ for $m\in \{-M,M\}$. The parametric modulation amplitude is then well approximated by~(see the supplement of \refcite{Dutt2019}):
\begin{equation}
  \label{eq:parametric-modulation-weights}
  v_{l}(\tau)=\frac{\Omega}{2} \frac{V(\tau)}{V_{\pi}} \sinc\pqty{l \frac{\phi}{2}},
\end{equation}
where \(V(\tau)\) is an applied voltage, \(V_{\pi}\) is the \(\pi\)-voltage of the EOM, and \(\phi=2\pi d/L\) is the angular length of the EOM.  If \(l\ll {1}/{\phi}\), the drive amplitude \(v_l(\tau=k/\Omega)\simeq v(\tau=k/\Omega)\) is approximately independent of \(l\) and is thus approximately equal to the dispersion relation \(\varepsilon(k)\) [\cref{eq:epsilon-approximate-dispersion}]. Alternatively, one can directly compensate for a small variation in the drive amplitude with $l$ by modifying the Fourier coefficients of $V(\tau)$ at the associated frequencies $l\Omega$. For typical values (taken from \refcite{Pellerin2024}), \(\Omega/2\pi = \SI{10}{\mega\hertz}\), \(\omega_{0}/2\pi\approx \SI{1.94e14}{\hertz}\) (\(\lambda_{0}=2\pi/k_0=\SI{1550}{\nano\meter}\), \(n_{0}\approx 1.5\)), it follows that \(\omega_{0}/\Omega\approx 10^{9}\)~\cite{Dutt2019,Yang2022a,Pellerin2024}.  This allows for the limit $\omega_0\gg M\Omega$ with $M\gg 1$, while neglecting optical dispersion that would lead to a frequency-dependent mode spacing. For example, in the low-dispersion optical fiber PM1550-XP (Thorlabs, USA)~\cite{thorlabsPM1550}, the relative deviation of the free spectral range is \(\Delta\Omega/\Omega<10^{-4}\) over \(N \sim 10^{4}\) modes.

\begin{figure}
  \centering
  \includegraphics[width=\columnwidth]{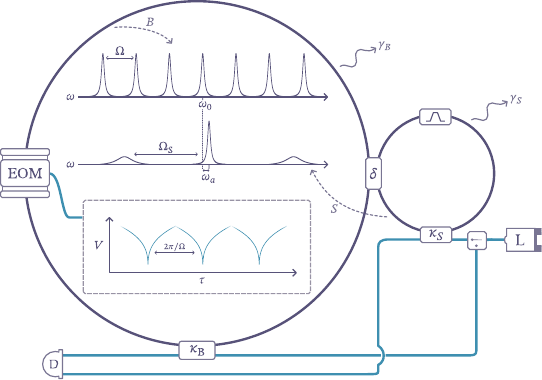}
  \caption{\label{fig:new_snowman} {A} fiber-loop setup that can be used to emulate dynamics of the Wigner-Weisskopf model. The bath is implemented in a big fiber loop with free spectral range \(\Omega/2\pi\). A small loop with free spectral range \(\Omega_{\sys}/2\pi\) supports a single long-lived mode with coherent-state amplitude $\alpha$ that emulates the unnormalized amplitude and phase for the excited state of a two-level system. The system and bath loops are coupled using a fiber coupler with strength \(\delta \ll \Omega\). The intrinsic photon loss rates of the system and bath loops are \(\gamma_{\sys}\) and \(\gamma_{\bath}\), respectively.  The small loop contains a filter with a bandwidth narrow enough for all but one mode at \(\omega_{0}+\omega_{a}\) to be heavily damped, ensuring that only one mode of the small loop participates appreciably in the dynamics. The big loop features an electro-optic modulator (EOM) to synthesize the bath spectral density. The big loop and the small loop are coupled to a transmission line using fiber couplers with coupling strengths \(\kappa_{\bath,\sys}\ll \delta\). A laser L can be switched between the small loop and the big loop. The output signal is detected with a photodiode D.}
\end{figure}

The proposed experimental setup (\cref{fig:new_snowman}) consists of two fiber loops, one big and one small, joined by a fiber coupler leading to a mode coupling \(\delta\).  The big loop (or bath loop), has a mode spacing \(\Omega\) and an intrinsic photon loss rate \(\gamma_{\bath}\). The smaller loop (system loop), has a mode spacing \(\Omega_{\sys}>\Omega\) and a loss rate \(\gamma_{\sys}\). The system loop hosts a single mode detuned by \(\omega_a\) from the bath-loop center frequency \(\omega_0\). A spectral filter is assumed to damp all modes in the system loop other than the one at frequency \(\omega_0+\omega_a\) with a loss rate \(\gamma'\gg \Omega\). In this limit, the dynamics of the system loop are approximately restricted to a single mode allowing us to consider an isolated single harmonic mode in the system loop coupled to many harmonic modes in the bath loop. Further, similar to the setup described in \cref{sec:meas-spectr-dens,fig:io_schematic}, a transmission line is assumed to be connected to one loop or the other in order to initialize the system by populating the relevant mode or to probe the field in the bath loop.

\subsection{Mapping to the Wigner-Weisskopf Model}
\label{sec:mapping-to-ww}

The Wigner-Weisskopf model describes a two-level atom interacting with a bath of electromagnetic field modes distributed in frequency with some density of states. The atom (system) has an excited state $\ket{e}$ and a ground state $\ket{g}$. We take the system Hamiltonian to be $H_\sys=\omega_a\ketbra{e}{e}$. An initial pure state of the system and bath consists of an arbitrary superposition of $\ket{e}$ and $\ket{g}$, and the vacuum for modes $k$ of the bath: $\ket{\psi(0)}=\psi_e(0)\ket{e,0}+\psi_g(0)\ket{g,0}$. This state will evolve under unitary evolution generated by $H$ [\cref{eq:system-bath-generic} with $L=\ketbra{g}{e}$] into
\begin{equation}\label{eq:psi_tau}
  \ket{\psi(\tau)} =\psi_g(0)\ket{g,0}+ \psi_e(\tau)\ket{e,0}+\sum_k\psi_k(\tau)\ket{g,k},
\end{equation}
where $\ket{g,k}=b_k^\dagger\ket{g,0}$. Given a dense spectrum of bath modes with a sufficiently weak coupling, the amplitude $\psi_e(\tau)$ may decay fully in time, but if no Markov approximation is made, $\psi_e(\tau)$ may show partial decay, revivals, or persistent oscillations~\cite{Cohen-Tannoudji1977,Gaveau1995,Krinner2018,Ricottone2020}.

The dynamics of the coefficients in \cref{eq:psi_tau} can be generated using the setup of system and bath loops discussed in \cref{sec:experimental-setup}. In particular, we consider the model Hamiltonian given in \cref{eq:final-cavity-bath} when \(\omega_{a}^{(2)}=0\). Since we only consider linear coupling, dynamics generated by the coupled system and bath loops maps coherent states to coherent states. In this case, the amplitudes $\psi_e(\tau)$ and $\left\{\psi_k(\tau)\right\}$, which are solutions of the time-dependent Schrödinger equation $\partial_\tau \ket{\psi(\tau)}=-i H\ket{\psi(\tau)}$, can be reproduced by the equations of motion for the coherent-state amplitudes of the fiber-loop modes
\begin{equation}
  \label{eq:classical-amplitude-transition}
  \beta_{k}(\tau)\equiv \ev{b_{k}(\tau)} \qq{and} \alpha(\tau)\equiv \ev{a(\tau)}.
\end{equation}
This is achieved in the limit of vanishing loss rates, $\kappa=\gamma_S=\gamma_B=0$, by identifying the coherent-state amplitudes as unnormalized versions of the probability amplitudes from \cref{eq:psi_tau} (see also \refcite{Yuan2016,Ozawa2016,Yuan2018,Dutt2019, bernazzani2024fluctuating} for a similar treatment in related models):
\begin{equation}
  \psi_e(\tau)=\alpha(\tau)/\mathcal{N},\quad \psi_k(\tau)=\beta_k(\tau)/\mathcal{N},
\end{equation}
where, for the initial condition chosen above [$\beta_k(0)=0$], the normalization factor is given by
\begin{equation}
  \mathcal{N}=\sqrt{|\psi_g(0)|^2+|\alpha(0)|^2}.
\end{equation}

In isolation from any external transmission lines ($\kappa=0$), but accounting for finite intrinsic losses, the coherent-state amplitudes for the bath loop and system loop are determined by the equations of motion
\begin{equation}
  \label{eq:coupled-amplitude-evolution}
  \begin{aligned}
    \dot{\alpha}    & = -\iu \omega_{a}\alpha  -\frac{\gamma_{\sys}}{2} \alpha -\iu \frac{\delta}{\sqrt{N}} \sum_{k} \beta_{k},  \\
    \dot{\beta}_{k} & = -\iu \varepsilon_{k} \beta_{k} - \frac{\gamma_{\bath}}{2} \beta_{k} -\iu \frac{\delta}{\sqrt{N}} \alpha,
  \end{aligned}
\end{equation}
where the bath energies $\varepsilon_{k}$ can be directly controlled through the parametric modulation amplitude $v_l(\tau)\simeq v(\tau)$ [\cref{eq:epsilon-approximate-dispersion}]. Integrating the equation of motion for
\(\beta_{k}\), for each $k$, and then substituting the solution into the equation for \(\alpha(\tau)\) gives (for $\gamma_\mathrm{S}=\gamma_\mathrm{B}=0$),
\begin{equation}
  \label{eq:excited-state-dynamics-int}
  \dot{\alpha}(\tau) = -\iu \omega_{a} \alpha(\tau) - \delta^{2}\int_{0}^{\tau}\dd{\tau'}{c}(\tau-\tau')\,\alpha(\tau'),
\end{equation}
with the bath correlation function $c(\tau)$ given by \cref{eq:bcf-stat}. \Cref{eq:excited-state-dynamics-int} demonstrates the history dependence of the dynamics of \(\alpha(\tau)\) through the convolution with the bath correlation function \(c(\tau)\); the present evolution of $\alpha(\tau)$ depends on its entire history, weighted by $c(\tau)$.

The observable of interest in the present work is the probability to remain in the excited state
\begin{equation}
  \label{eq:excited-state-population-probability}
  p_{e}(\tau) \equiv \abss{\psi_e(\tau)}=\frac{{\abss{\alpha(\tau)}}}{\mathcal{N}^{2}}.
\end{equation}
In the thermodynamic limit ($N\to\infty$) and when choosing the energies \(\varepsilon_{k}\) according to the power-law dispersion relation given in \cref{eq:dispersion-powerlaw}, this probability has a long-time average that either vanishes or remains finite depending on the spectral exponent \(s\) of the bath~\cite{Palma1997,Ricottone2020}:
\begin{equation}
  \label{eq:population-decay-conditional}
  {\peb}\equiv \lim_{\tau\to\infty} \frac{1}{\tau} \int_{0}^{\tau}\dd{t}  p_{e}(t)=
  \begin{cases}
    0,          & 0\leq s \leq 1 \\
    \peb\neq 0, & s > 1          \\
  \end{cases}.
\end{equation}
In the rest of this section, we will derive closed-form expressions for $\peb$ accounting for finite-size effects.

Finding an exact closed-form solution to the integrodifferential equation, \cref{eq:excited-state-dynamics-int}, is simple only in special cases, but the equation can be recast as an algebraic expression by invoking the Laplace transform
\begin{equation}
  \alpha(z) = \int_0^\infty \dd{\tau} e^{-z\tau}\alpha(\tau).
\end{equation}
This leads directly to the solution
\begin{equation}
  \label{eq:laptransformed-solution}
  \alpha(z) = \frac{\alpha_0}{z+i\omega_a+\delta^2 c(z)},
\end{equation}
where $\alpha_0=\alpha(\tau=0)$ is the initial condition. The time-domain quantity $\alpha(\tau)$ can then be found, in principle, by performing the Bromwich inversion integral
\begin{equation}
  \alpha(\tau) = \lim_{\eta\to 0^+}\int_{-i\infty+\eta}^{i\infty+\eta}\frac{\dd z}{2\pi i}e^{z\tau}\alpha(z).
\end{equation}
For a finite bath with discrete levels, the inversion integral can be evaluated as a sum over residues at the poles of $\alpha(z)$. In the continuum limit, the discrete sum over residues associated with the bath is replaced by a continuous branch-cut integral, but there may also be discrete poles along the imaginary axis (see also \cref{sec:deta-wign-weissk} for a detailed discussion). For sufficiently weak coupling $\delta$ and for $\omega_a<0$ so that the bare-atom frequency lies below the continuum band, there will be an isolated pole at $z\simeq -i\omega_a$ corresponding to the atom frequency, but a finite coupling to the bath will modify this frequency by the Lamb shift $\lamb$ \cite{Weisskopf1930,Cohen-Tannoudji1977,Breuer2002,Fick1990} to a new renormalized atom frequency $\omega_a\to\omega_a^*$, where
\begin{equation}
  \label{eq:astar-def}
  \omega_a^*=\omega_a+\lamb.
\end{equation}
The Lamb shift can be found by noting that:
\begin{equation}\label{eq:CorrelationRealImag}
  c(z=-i\omega+0^+)=J(\omega)+i\mathcal{P}\int \dd\omega' \frac{D(\omega')}{\omega-\omega'},
\end{equation}
where $\mathcal{P}$ indicates a principal value integral. The real part of \cref{eq:CorrelationRealImag} gives the spectral density $J(\omega)$, describing decay into the bath in the continuum limit, and the imaginary part results in frequency shifts. In particular, the Lamb shift is given by the self-consistent solution to
\begin{equation}
  \label{eq:lambformula}
  \lamb = \delta^2\mathcal{P}\int \dd\omega' \frac{D(\omega')}{\omega_a+\lamb-\omega'}.
\end{equation}
When $\omega_a^*=\omega_a+\lamb<0$, there will be at least one isolated pole on the imaginary axis. The long-time average of $\alpha'(\tau)=e^{i\omega_a^*\tau}\alpha(\tau)$ may then result in a non-decaying contribution given by the residue
\begin{equation}
  \overline{\alpha}=\lim_{\eta\to 0^+} \eta \int_0^\infty\dd\tau e^{-\eta \tau}\alpha'(\tau)=\lim_{z\to -i\omega_a^*}(z+i\omega_a^*)\alpha(z).
\end{equation}
This gives (for $\omega_a^*<0$):
\begin{equation}
  \label{eq:alphabardef}
  \overline{\alpha}=\frac{\alpha_0}{1+\delta^2 c^\prime(z=-i\omega_a^*)},
\end{equation}
where
\begin{equation}
  \label{eq:residue}
  c^\prime(z=-i\omega_a^*) = \left.\frac{\dd c(z)}{\dd z}\right|_{z=-i\omega_a^*}=\int \dd\omega \frac{D(\omega)}{(\omega-\omega_a^*)^2}.\end{equation}
The expressions above apply generally, either for a discrete and finite bath (finite $N$), where $D(\omega)$ is expressed in terms of a sum over Dirac delta functions, or for an infinite bath in the continuum limit ($N\to\infty$), where $D(\omega)$ is a continuous function of frequency.

The continuum expression for $D(\omega)\propto \omega^s$ from \cref{eq:power-law-spectral-density} vanishes as $\omega\to 0^+$, as required for a bath with a spectrum that is bounded from below. If we select the bare atom frequency $\omega_a$ so that the renormalized atom frequency $\omega_a^*=\omega_a+\lamb$ approaches the edge of the continuum band from below ($\omega_a^*\to 0^-$), then inserting the expression for $D(\omega)$ from \cref{eq:power-law-spectral-density} into the expressions above gives
\begin{equation}\label{eq:alpha_piecewise}
  \lim_{\omega_a^*\to 0^-}\overline{\alpha}=\begin{cases}
    0,                                                                              & 0 < s \leq 1, \\
    \frac{\alpha_0}{1+\frac{s+1}{s-1}\left(\frac{\delta}{\omega_c}\right)^2}\neq 0, & s > 1.        \\
  \end{cases}
\end{equation}
In the same limit, the Lamb shift is given by
\begin{equation}
  \label{eq:lam}
  \lamb = \lim_{\omega_a^*\to 0^-}\delta^2\mathcal{P}\int \dd\omega' \frac{D(\omega')}{\omega_a^*-\omega'}=-\frac{\delta^{2}}{\omega_{c}} \frac{1+s}{s}.
\end{equation}

\begin{figure}[tp]
  \centering
  \includegraphics{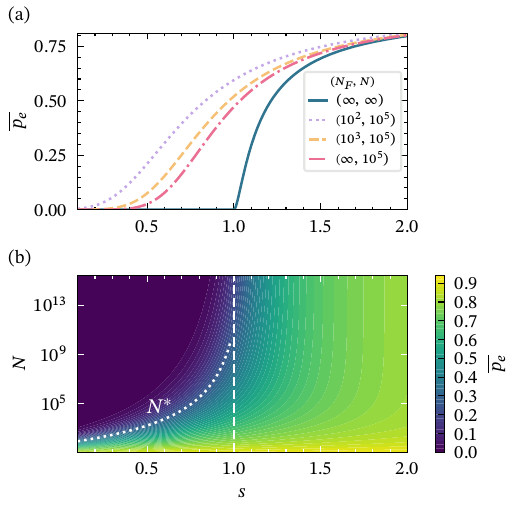}
  \caption{\label{fig:exact_phasemap} The long-time averaged excited-state probability \(\peb\simeq{|{\overline{\alpha}}|}^{2} /|\alpha_0|^2 \), with \(\overline{\alpha}\) evaluated from \cref{eq:alpha_bar_finite_NF} for (a) various \((N_F, N)\), where $N$ sets the number of bath modes and $N_F$ scales with the number of Fourier coefficients used to approximate the dispersion $\bar{\varepsilon}_{sk}$ [\cref{eq:filtered_bath_energies}], and (b) for $N_F\to\infty$, where the parameter \(N^*\) [dotted line, \cref{eq:N-star_definition}] signals the transition to a power-law suppression of $\peb$ with increasing $N$, and where the dashed vertical line marks the transition at \(s=1\) from complete decay to partial decay in the continuum limit, \((N,N_F)\to(\infty,\infty)\). The special limiting case \((N,N_F)\to(\infty,\infty)\) [solid blue curve in subfigure (a)] is given by \cref{eq:excited_state_prob_asymp}. These plots were made assuming \({\delta} / {\omega_{c}} = 1 / 5\), ensuring that the results are in the weak-coupling regime for $s\gtrsim 0.04$ [cf. \cref{eq:weak_coupling_suppression}].}
\end{figure}

As described above, in the continuum limit (\(N\to\infty\)) there are only two types of contribution to \(\alpha(\tau)\): the branch-cut integral and residues at isolated poles in $\alpha(z)$. The branch-cut contribution decays with time, so that in the long-time limit there remain only the contributions from two isolated poles on the imaginary axis; one pole at $z\simeq -i\omega_a^*$, as discussed above, and the other approaching \(z=-\iu \omega_{c}\) in the weak-coupling limit. The long-time time-averaged excited-state probability then becomes
\begin{equation}\label{eq:p_e_avg}
  \overline{p_e}=\lim_{T\to\infty}\frac{1}{T}\int_0^T \dd\tau |\psi_e(\tau)|^2= \frac{{|\overline{\alpha}|}^{2}}{\mathcal{N}^{2}} + \mathcal{O}\pqty{\eu^{-\Lambda}},
\end{equation}
where the correction due to the pole at $z\simeq -i\omega_c$ is exponentially suppressed ($\Lambda\gg 1$) in the weak-coupling regime,
\begin{equation}
  \label{eq:weak_coupling_suppression}
  \pqty{\frac{\delta}{\omega_{c}}}^{2} \ll \frac{s}{1+s},
\end{equation}
provided the renormalized atom frequency matches the continuum band edge, $\omega_a^*=0$. The right-hand side of \cref{eq:weak_coupling_suppression} is an interpolating function that recovers the correct weak-coupling conditions for \(s\ll 1\) [\((\delta/\omega_c)^2\ll s\)] and for \(s\simeq 1\) [\((\delta/\omega_c)^2\ll 1/2\)]. It therefore applies for the typical range of $s$ considered here $0\le s\lesssim 2$ (see \cref{sec:deta-wign-weissk} for details). If the atom is initially prepared in the excited state [\(\psi_{g}(0)=0\)], then substituting the expression for \(\overline{\alpha}\) given in \cref{eq:alpha_piecewise} into the expression for \(\peb\) in \cref{eq:p_e_avg} gives (neglecting exponentially small corrections),
\begin{equation}
  \label{eq:excited_state_prob_asymp}
    \overline{p_e}\simeq\begin{cases}
    0,                                                                              & 0 < s \leq 1, \\
    {\frac{1}{\bqty{1+\frac{s+1}{s-1}\left(\frac{\delta}{\omega_c}\right)^2}^{2}}}  & s > 1.        \\
  \end{cases}
\end{equation}
\Cref{eq:excited_state_prob_asymp} [shown as the blue solid line in \cref{fig:exact_phasemap}(a)] describes the localization transition from partial decay for a superohmic spectral density ($s>1$) to complete decay for an ohmic or subohmic spectral density ($s\le 1$) in the weak-coupling regime $\delta\ll \omega_c$ [cf.~\cref{eq:weak_coupling_suppression}] and for $\omega_a^*=0$ \cite{Cohen-Tannoudji1977,Gaveau1995,Krinner2018,Ricottone2020}.

Outside of the weak-coupling regime, two poles in $\alpha(z)$ located on the imaginary axis will contribute significant weight: one at $z\to -i\omega_a$ as $\delta\to 0$, and a second one that emerges on the opposite side of the continuum band as $\delta$ is increased to a value $\delta > \omega_{c}$. In this strong-coupling regime, the combination of the two isolated pole contributions leads to a non-decaying beating signal in $p_e(\tau)\propto |\alpha(\tau)|^2$~\cite{khaetskii2002electron,Taylor2003}, with
\begin{equation}
  \label{eq:coherent-osci}
  \alpha(\tau) \simeq \alpha(0) \cos(\delta\tau).
\end{equation}
In the strong-coupling regime, this beating occurs completely independent of the choice of spectral exponent \(s\).
In \refcite{Taylor2003} these strong-coupling effects were suggested as a method to store and retrieve information in a mesoscopic quantum memory. Although the focus of \refcite{khaetskii2002electron,Taylor2003} was the problem of an electron spin interacting with a polarized nuclear-spin bath, the formal problem studied there is identical to the Wigner-Weisskopf problem described here, with the two states of an electron spin replacing the two atomic states and with spin flips in the nuclear-spin system replacing excitations of the electromagnetic field for Wigner-Weisskopf dynamics. Very similar strong-coupling effects have also been theoretically analyzed and experimentally measured for an ensemble of Rydberg atoms under strong driving \cite{Covolo2025}.

\subsection{Observing the localization transition}
\label{sec:constraints}
The goal of this subsection is to explain potential experimental limitations in realizing the localization transition from complete decay (for $s\le 1$) to partial decay (for $s>1$). The focus will be on experimentally emulating this transition in the fiber-loop cavity described in \cref{sec:experimental-setup,sec:mapping-to-ww}, but several of the issues described in this subsection will be relevant to other experimental implementations as well.  Finite-size effects are studied in \cref{sec:finite-size-effects}. Limitations due to a finite bandwidth of the parametric drive are studied in \cref{sec:effects-finite-drive}. The role of losses is considered in \cref{sec:losses}. Numerical simulations are presented in \cref{sec:simulations} to demonstrate what  an experiment with realistic parameters might be able to measure.

\subsubsection{Finite-size effects}
\label{sec:finite-size-effects}
In an experimental implementation of the localization transition described above, finite-size effects are controlled by two parameters: (i) a finite number of nonzero Fourier coefficients $\propto N_F$ used to generate time-dependent parametric modulation leading to the approximate bath dispersion [\cref{eq:filtered_bath_energies}] and (ii) a finite number of discrete bath modes $N$. If we formally take $N_F\to\infty$, resulting in the idealized bath dispersion, then there can still be finite-size effects for a large but finite \(N\gg 1\).

To obtain a large-\(N\) expression for \(\peb\) in the weak-coupling regime, we neglect the exponentially small correction \(\sim \mathcal{O}(e^{-\Lambda})\) in \cref{eq:p_e_avg}, set \(\psi_{g}(0)=0\) (so that \(\mathcal{N}=\abs{\alpha_{0}}\)), and evaluate \(\overline{\alpha}\) from \cref{eq:alphabardef} for a discrete bath with the spectral density given in \cref{eq:josephson-density-function}:
\begin{equation}\label{eq:alpha_bar_finite_NF}
  \overline{\alpha} = \lim_{\omega_a^*\to 0^-}\frac{\alpha_0}{1+\frac{\delta^2}{N}\sum_k \frac{1}{\pqty{\esk_{sk}-\omega_a^*}^2}}.
\end{equation}
Above, we have used \(\varepsilon_{k}=\esk_{sk} = \sum_{l=-N_F}^{N_F} t_{sl} e^{-ikl}\) [\cref{eq:filtered_bath_energies}] with the drive Fourier coefficients \(t_{sl}\) defined in \cref{eq:hop-amp}.  The limit $\omega_a^*\to 0^-$ exists for any finite $N_F$ since $\esk_{sk}$ remains nonzero for all $k$ at any finite $N_F$ [see \cref{subfig:lowpassdisp} for examples with $N_F=100$]. However, if the limit $N_F\to\infty$ is taken first, then the $k=0$ contribution to the denominator of \cref{eq:alpha_bar_finite_NF} would diverge since the idealized dispersion $\varepsilon_s(k)\propto |k|^{1/(s+1)}$ vanishes at $k=0$. In the thermodynamic limit ($N\to\infty$), the set of $k$-states becomes continuous and the contribution from $k=0$ becomes measure zero, so taking the thermodynamic limit ($N\to\infty$) first, then taking the infinite-bandwith limit ($N_F\to\infty$) leads to a meaningful result. This same result can be found starting from \cref{eq:alpha_bar_finite_NF} simply by omitting the $k=0$ contribution and interchanging the order of limits (taking $N_F\to\infty$, then $N\gg 1$) to isolate the leading large-$N$ behavior.

To obtain an analytic expression under the conditions described above, we convert the sum in the denominator of \cref{eq:alpha_bar_finite_NF} to an integral via the Euler-Maclaurin formula, neglecting corrections that are higher-order in $1/N$. This gives
\begin{equation}
\frac{\delta^2}{N}\sum_{k (\ne 0)} \frac{1}{\varepsilon_k^2}\simeq \left(\frac{\delta}{\omega_c}\right)^2\left(\frac{2}{N}\right)^\epsilon\int_1^{\frac{N-1}{2}}\dd m m^{\epsilon-1},
\end{equation}
where we have used the infinite-bandwidth ($N_F\to\infty$) form of the dispersion, $\varepsilon_k=\varepsilon(k)=\omega_c(|k|/\pi)^{1/(s+1)}$, with $k=2\pi m/N$, and with $m$ ranging from $-M$ to $M=(N-1)/2$. In addition, for convenience we have introduced the symbol
\begin{equation}
\epsilon=\frac{s-1}{s+1}.
\end{equation}
Performing the integral and substituting the result back into the formula for $\peb = |\overline{\alpha}/\alpha_0|^2$, then neglecting corrections that are higher-order in $1/N\ll 1$ gives:
\begin{equation}
  \label{eq:pebarasymp}
  \peb \simeq \frac{1}{\bqty{1+ \pqty{\frac{\delta}{\omega_{c}}}^{2}\frac{1}{\epsilon}\pqty{1-\bqty{\frac{2}{N}}^{\epsilon}}}^{2}};\quad (N\gg 1).
\end{equation}
This expression recovers the result reported in \cref{eq:excited_state_prob_asymp} in the limit $N\to\infty$, but it also describes the leading finite-$N$ corrections.

In the subohmic regime $(s<1,\epsilon<0)$, we find the large-$N$ behavior,
\begin{equation}\label{eq:pe_subohmic}
    \peb \simeq \left(\frac{\omega_c}{\delta}\right)^4|\epsilon|^2\left(\frac{N}{2}\right)^{-2|\epsilon|};\quad N\gg N^*\gg 1,
\end{equation}
where
\begin{equation}\label{eq:N-star_definition}
N^* = 2\bqty{1+|\epsilon|\left(\frac{\omega_c}{\delta}\right)^2}^{1/|\epsilon|}.
\end{equation}
For the Ohmic limit $(s\to 1,\epsilon\to0)$,
\begin{equation}\label{eq:pe_ohmic}
\peb \simeq \left(\frac{\omega_c}{\delta}\right)^4\frac{1}{\ln^2(N/2)};\quad N\gg 2 e^{\left(\omega_c/\delta\right)^2},
\end{equation}
and in the superohmic regime $(s>1,\epsilon>0)$,
\begin{equation}\label{eq:pe_superohmic}
\peb \simeq \frac{\epsilon^2}{\left[\epsilon+\left(\frac{\delta}{\omega_c}\right)^2\right]^2};\quad N\gg 1,
\end{equation}
which recovers \cref{eq:excited_state_prob_asymp} with $\epsilon = (s-1)/(s+1)$.

In the subohmic regime and for $|\epsilon|\sim O(1)$, $\peb\propto 1/N^{2|\epsilon|}$ is suppressed for increasing $N\gg N^* \propto (\omega_c/\delta)^{2/|\epsilon|}$, but this suppression will require a larger number of modes $N$ for a weaker coupling (leading to a larger ratio $\omega_c/\delta$) and the requirement for a large number of modes becomes progressively more stringent closer to the Ohmic limit, $\epsilon\simeq 0$. In the Ohmic limit $\epsilon\simeq 0$, the number of bath modes $N$ required to accurately describe the thermodynamic limit becomes exponential in the ratio $\omega_c/\delta$ [\cref{eq:pe_ohmic}]. This could make experimentally achieving results similar to the thermodynamic limit challenging in the weak-coupling regime $\delta/\omega_c\ll 1$ if, in addition, only a modest number of bath modes $N$ is accessible. As mentioned around \cref{eq:coherent-osci}, above, the strong-coupling limit $\delta/\omega_c\gtrsim 1$ leads to a qualitatively different dynamics. \Cref{fig:exact_phasemap} shows $\peb$ for a compromise value $\delta/\omega_c=1/5$ that is in the weak-coupling regime, but is not so deep into that regime to require an unimaginable number of bath modes $N$. For $\delta/\omega_c=1/5$, we would still require $N\gg 2e^{(\omega_c/\delta)^2}\sim 10^{11}$ to find a sharp transition at $s=1$, as shown in \cref{fig:exact_phasemap}(b). As described in \cref{sec:experimental-setup} above, $N\sim 10^4$ modes can be accessed in fiber-loop cavities, but reaching $N\sim 10^{11}$ in this particular setup is likely not possible because dispersion leads eventually to an unequal mode spacing. Nevertheless, as described above, there will be a strong algebraic suppression of $\peb$ at some crossover point \(s=s^*<1\) provided $N>N^*$, so the phenomenology of the phase transition is retained even for a finite bath and a finite coupling. For example, at $N^*\sim 10^4$, and $\delta/\omega_c=0.2$, the crossover from near-complete to partial decay occurs at $s^*\simeq 0.64$ [\cref{fig:exact_phasemap}(b)]. Moreover, as we demonstrate numerically in \cref{sec:simulations}, below (see also \cref{fig:comp_ideal_v_real}) and justify in \cref{sec:fidelity-finite-size}, although the long-time dynamics leading to $\peb$ may be difficult to realize experimentally, the short-time dynamics of the ideal model ($N \to \infty, N_F \to \infty$) can be accurately realized with much more modest resources.

\subsubsection{Effects of a finite drive bandwidth}
\label{sec:effects-finite-drive}
As illustrated in \cref{fig:dispersion-plot}, the dispersion relation \(\esk_{sk}= \sum_{l=-N_F}^{N_F} t_{sl} e^{-ikl}\) realized in an experiment using a finite number of drive Fourier coefficients \(N_{F}\) deviates from the ideal dispersion \(\varepsilon_{s}(k)\propto |k|^{1/(s+1)}\) [\cref{eq:dispersion-powerlaw}] only in a small region of the Brillouin zone, \(\abs{k}\lesssim \Delta/N_{F}\) for some constant $\Delta\ll N_F$. Since that range of $k$ becomes measure zero in the limit $N_F\to\infty$, we can convert the sum in \cref{eq:alpha_bar_finite_NF} to an integral in the limit $N\to\infty$ and then restrict the integration region to $k>\pi\Delta/N_F$, replacing $\esk_{sk}\simeq \varepsilon_{s}(k)$ for $k>\pi\Delta/N_F$. This approximation gives
\begin{equation}
\frac{\delta^2}{N}\sum_{k}\frac{1}{\esk_{sk}^2}\simeq \left(\frac{\delta}{\omega_c}\right)^2\int_{\pi \Delta/N_F}^\pi \frac{dk}{\pi}\left(\frac{k}{\pi}\right)^{\epsilon-1}.
\end{equation}
In the Ohmic limit $(s\to1,\epsilon\to 0)$, this gives
\begin{equation}
\frac{\delta^2}{N}\sum_{k}\frac{1}{\esk_{sk}^2}\simeq \left(\frac{\delta}{\omega_c}\right)^2\ln\left(\frac{N_F}{\Delta}\right).
\end{equation}
Substituting this result into the expression for $\peb$ leads to the large-$N_F$ expression
\begin{equation}
  \label{eq:peb-nf-scale}
\peb \simeq \left(\frac{\omega_c}{\delta}\right)^4\frac{1}{\ln^2 N_F};\quad N_F\gg e^{\left(\omega_c/\delta\right)^2}\gg \Delta,
\end{equation}
for some $\Delta=O(1)$. Suppressing $\peb$ to zero in the Ohmic limit ($s=1$) thus requires $N_F$ to be exponentially large in $\omega_c/\delta$, similar to the case for $N\gg 1$ at $N_F\to\infty$. However, as in the finite-$N$ case, there will still be a strong algebraic suppression of $\peb$ with increasing $N_F$ for some $s<s^*<1$. Furthermore, as noted in \cref{sec:finite-size-effects} above, if we instead consider finite-time dynamics, we can accurately reproduce the ideal dynamics for a more modest number of Fourier coefficients \(N_{F}\).

\subsubsection{Photon loss}
\label{sec:losses}

Realistic cavities are subject to photon loss. To quantify this effect, we consider the transformed (damping-compensated) amplitudes
\begin{equation}
  \label{eq:transformed-amplitude-exponentiation}
  \tilde{\alpha}(\tau)\equiv \alpha(\tau) \eu^{\frac{\gamma_{\sys}}{2}\tau}\qq{and} \tilde{\beta}_{k}(\tau)\equiv \beta_{k}(\tau)\eu^{\frac{\gamma_{\bath}}{2}\tau}.
\end{equation}
The equation of motion for $\tilde{\alpha}(\tau)$ can be obtained from the equation for $\alpha(\tau)$ [\cref{eq:excited-state-dynamics-int}] by replacing \(c(\tau)\) with an effective bath correlation function
\begin{equation}
  \label{eq:effective_bcf}
  c_{\text{eff}}(\tau)= c(\tau)\eu^{-\frac{\Delta\gamma}{2}\abs{\tau}},
\end{equation}
where we have introduced the damping asymmetry \(\Delta\gamma = \gamma_{\bath}-\gamma_{\sys}\). If the system- and bath-cavity decay rates are balanced ($\gamma_S=\gamma_B=\gamma$, $\Delta\gamma=0$), then the dynamics of $\tilde{\alpha}(\tau)$ will accurately describe the dynamics $\alpha(\tau)$ in the absence of damping. In an experimental setting, where measurements give noisy estimates of $\alpha(\tau)$ and $\beta(\tau)$, the transformation in \cref{eq:transformed-amplitude-exponentiation} is only meaningful if the amplitude of the noise is smaller than the amplitude of the exponentially damped signal. To ensure this is the case, it may be advantageous to restrict the observation time $\tau$ to be shorter than some maximum time $\tau_{\max}$ satisfying
\begin{equation}
  \label{eq:timescale-conditions}
  \tau<\tau_{\max} \lesssim \min\left(\frac{1}{\gamma_{\sys}},\frac{1}{\gamma_{\bath}}\right).
\end{equation}
 Additionally, the effect of a nonzero damping asymmetry \(\Delta\gamma\) can only be neglected up to a finite timescale. We find a sufficient condition to neglect the effects of a finite \(\Delta\gamma\) is:
\begin{equation}
  \label{eq:damping_asym_timescale}
  \tau\ll\tau_{\max}= \pqty{\frac{6}{\delta^{2}\abs{\Delta\gamma}}}^{\frac{1}{3}}.
\end{equation}
For details on the derivation of \cref{eq:damping_asym_timescale}, see the text surrounding \cref{eq:final-bound}, below.

Finally, even in the limit of vanishing damping asymmetry, $\Delta\gamma\simeq 0$, the photon loss rate $\gamma=\gamma_S\simeq \gamma_B$ must be sufficiently small to realize a quasi steady-state in $p_e(\tau)e^{\gamma\tau}$ before the signal is fully damped. This condition is crucial to observe the localization transition. To estimate the timescale required to reach a quasi steady-state, we consider a flat (frequency-independent) spectral density ($s=0$) with the effective atom frequency $\omega_a^*$ lying within the continuum band, $\omega_a^* \in (0,\omega_c)$. In the continuum limit ($N\to \infty, N_F \to \infty$) and assuming weak decay ($\Gamma \ll \omega_c$), the Markov approximation yields an exponential decay for the excited-state probability, $p_e(\tau) \approx e^{-\Gamma\tau}$. This procedure (\refcite[chapter 13.1]{Fick1990}) yields the decay rate:
\begin{equation}
\Gamma = 2\pi \delta^2 D(\omega_a^*)=\frac{2\pi\delta^2}{\omega_c},
\end{equation}
a result familiar from Fermi's golden rule. This rate sets an approximate time scale $\simeq 1/\Gamma$ for transients to die off before $p_e(\tau)$ reaches a saturation value. For an experiment to accurately describe the saturation value on a time scale long compared to this initial transient decay time, but short compared to the time for photon loss, we have \(\tau_{\max}\sim {1} /{\Gamma}\lesssim 1/\gamma\). This condition then constrains the loss rate:
\begin{equation}
  \label{eq:decay-condition-constraints}
  \gamma \lesssim \Gamma\implies \gamma \lesssim \frac{2\pi \delta^{2}}{\omega_{c}}.
\end{equation}
For the parameters given in \cref{tab:simuparam} (used to produce \cref{fig:comp_ideal_v_real}), we have \(1 / \Gamma \approx \SI{0.6}{\micro\second}\), which can be shorter than a typical decay time in fiber loops. For example, a lifetime \(1 / \gamma \simeq \SI{5}{\micro\second}\) can be inferred for the fiber-loop experiments of \refcite{Pellerin2024}, based on the resonance full width at half maximum, $\Delta\nu=\gamma/\pi$, reported in the supplemental material of that reference.

\subsubsection{Numerical simulations}
\label{sec:simulations}
\begin{figure}[ht]
\includegraphics{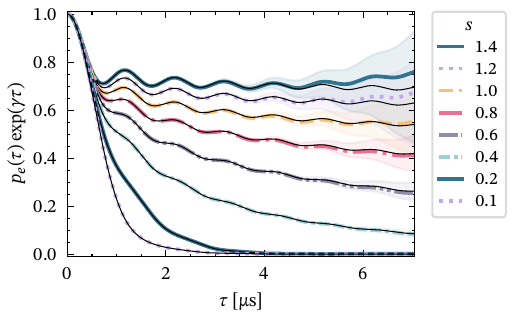}
  \caption{\label{fig:comp_ideal_v_real} Comparison of expected near-ideal dynamics (thin black solid lines) with dynamics that can be realized in a fiber-loop setup (thick colored lines), obtained by integrating the equations of motion given in \cref{eq:coupled-amplitude-evolution} using the parameters given in \cref{tab:simuparam}. The number of bath modes and Fourier coefficients used were $N$ and $N_F$, respectively, for the coloured lines and $N'=10N$ and $N_F'\to\infty$ for the black lines. See the main text for a description of the error funnels and for further details.}
\end{figure}

\Cref{fig:comp_ideal_v_real} shows the loss-compensated excited-state probability \(p_{e}(\tau)e^{\gamma\tau}\) as a function of time for various spectral exponents \(s\). The transition from complete decay (at $s=0.1$) to partial decay (at $s=1.4$) is clearly discernible. Thin black solid lines in \cref{fig:comp_ideal_v_real} show the dynamics for a reference set of parameters that we expect to accurately reproduce the thermodynamic limit. In particular, these curves were produced by numerically integrating the equations of motion given in \cref{eq:coupled-amplitude-evolution} (with \(\gamma_{\sys}=\gamma_{\bath}=\gamma\), i.e.~\(\Delta\gamma=0\)) using the exact continuum spectral density (\(N_{F}=\infty\)) and a bath of \(N'=2\times 10^{5}\) modes. For comparison, thick colored lines show the dynamics that can be obtained for more modest parameters that can be achieved in an optical-fiber-loop experiment, \(N_{F}=50,\,N=2\times 10^4\). See \cref{tab:simuparam} for a complete list of parameters. Funnels surrounding the colored lines correspond to an error bound [given in \cref{eq:pe_deviation_bound}, below] that will be discussed in \cref{sec:fidelity-finite-size}. The error funnels and simulations exclude additional corrections to the rotating-wave approximation due to the nonstationary phase $\phi_k(\tau)$. These corrections are suppressed in the small parameter $\Omega/\omega_c\ll 1$ (see \cref{sec:rotat-wave-appr,sec:rotat-wave-appr-1}).

\bgroup
\begin{table}[ht]
  \centering
  \begin{tabular}{c|c}
    Parameter           & Value                 \\
    \hline
    \hline
    \(N_{F}\)           & \(50\)                \\
    \(N\)               & \(2\times 10^{4}\)    \\
    \(\omega_{c}/2\pi\) & \SI{1}{\mega\hertz}   \\
    \(\delta\)          & \(\omega_{c}/5\)      \\
    \(\Omega/2\pi\)     & \SI{9.8}{\mega\hertz} \\
    \(N'\)              & \(2\times 10^{5}\)    \\
    \(N_F'\)            & \(\infty\)\\
    \(\gamma\) & $1/ (\SI{5}{\micro\second})$
  \end{tabular}
  \caption{Parameter choices used to obtain the simulation results in \cref{fig:comp_ideal_v_real}. With the exception of $N'$ and $N_F'$, these parameters are expected to be realistic for the setup based on fiber-loop cavities discussed in \cref{sec:experimental-setup}.}
  \label{tab:simuparam}
\end{table}
\egroup
To synthesize an approximate power-law spectral density \(J(\omega)\propto \omega^{s}\), the bath-mode energies were sampled from the filtered (Fourier-truncated) dispersion relation \(  \esk_{sk} = \sum_{l=-N_F}^{N_F} t_{sl} e^{-ikl} \) that recovers the ideal dispersion relation \(\varepsilon(k)=\omega_{c} \abs{k / \pi}^{1 / (1+s)}\) in the limit \(N_F\to\infty\). The Fourier coefficients $t_{sl}$ are given in closed form in \cref{sec:dispersion}, \cref{eq:hypergeometric-coefficients-formula}.
The Lamb shift has been compensated by choosing \(\omega_{a}=\omega_{a}^{*}-\lamb=-\lamb\) (\(\omega_{a}^{*}\approx 0\)) using the continuum-limit formula for the Lamb shift  [\cref{eq:lam}], resulting in \(\omega_{a} = ({\delta^{2}} / {\omega_{c}}) \pqty{1+s} / {s}\).

In the thermodynamic limit, the long-time behavior is given (for \(0\le s\le 2\)) by
\begin{equation}
  \label{eq:aysmp-fit}
  e^{\gamma_S\tau}p_e(\tau)\simeq \peb+A/\tau^{|s-1|};\quad (\tau\to\infty).
\end{equation}
 See \cref{sec:some-notes-therm} for details. Experimentally, it may be possible to extract \(\peb\) by fitting the measured data to \cref{eq:aysmp-fit} when $\omega_a^*\simeq 0$ for times $\tau$ that are sufficiently long to justify the asymptotic scaling, but still sufficiently short to avoid a significant deviation from ideal behavior (due to finite $N$, $N_F$, or to the damping asymmetry $\Delta\gamma\ne 0$). Alternatively, if the asymptotic regime is not reached, \(\peb\) can be extracted by time-averaging \(e^{\gamma_S\tau}p_{e}(\tau)\) over some interval that is long compared to the initial transient decay time \(\sim 1 / \Gamma\), but short compared to the time at which the funnels diverge. Both methods will eventually converge to \(\peb\) for a sufficiently large number of drive Fourier coefficients \(\propto N_{F}\) and bath size \(N\), for a sufficiently small loss rate and damping asymmetry \(\gamma,\Delta\gamma\), and for a sufficiently small nonstationary phase $\phi_k(\tau)$.
A set of bounds on deviations that can be induced through each of these sources of imperfection is derived in the following section.

\section{Bounds on emulating system-bath models}
\label{sec:fidelity-finite-size}
We now provide a rigorous quantitative assessment of the systematic error in emulating the dynamics of a system-bath model with a realistic physical apparatus. To do this, we make use of a non-perturbative analytical bound that limits the deviation of a system observable [e.g.~$p_e(\tau)$] that can be physically realized under imperfect conditions, relative to the dynamics induced by an idealized model. This bound is related to the time-integrated deviation in the bath correlation function for physical dynamics, relative to some idealized target bath correlation function. We then apply this general theory to the Wigner-Weisskopf model, where the observable of interest is the damping-compensated excited-state probability $e^{\gamma_\sys \tau}p_e(\tau)$. In this context, we derive a precise bound on the deviation of the bath correlation function in terms of key experimental parameters: the number of drive Fourier coefficients $\propto N_F$, the damping asymmetry $\Delta\gamma$, the non-stationary phase $\phi_k(\tau)$, and the bath size $N$.

\subsection{General bound}
\label{sec:fundamental-bound}
To quantify the degree to which a system-bath model can be emulated with a parametrically driven cavity, we compare two models of the type introduced in \cref{sec:open-quantum-systems}: an ideal \emph{target model} characterized by the stationary correlation function \(c_{0}(\tau,\tau')=c_{0}(\tau-\tau')\), and the \emph{physical model} realized in the experiment, with correlation function \(c(\tau,\tau')\). These models are assumed to be identical except for their bath realization; the system Hamiltonian \(H_{\sys}\) and system-bath coupling operator \(L\) are assumed to be identical in the two cases.

Our objective is to bound the deviation
\begin{eqnarray}
  \label{eq:obsdiff}
  \Delta \ev{O(\tau)}\equiv\abs{\ev{O(\tau)}-\ev{{O(\tau)}}_{0}},
\end{eqnarray}
where $\ev{O(\tau)}$ is the expectation value of a system observable \(O\) evolving under the physical model and where $\ev{{O(\tau)}}_{0}$ is the expectation value arising from the target model.
For any observable bounded by \(\mathcal{N}_{O}\) (i.e., ${\ev{O(\tau)}} \leq \mathcal{N}_O$), the deviation $\Delta \ev{O(\tau)}$ is limited by \cite{Huang2024}:
\begin{equation}
  \label{eq:observable-deviation-bound-lo}
  \begin{aligned}
    \Delta\ev{O(\tau)} & \leq
                 \mathcal{N}_{O}\bqty{\eu^{{4\delta^{2}\mathcal{N}_L^{2}}\int_{0}^{\tau}\dd{\tau_{1}}\int_{0}^{\tau_{1}}\dd{\tau_{2}}\abs{\Delta c(\tau_{1},\tau_{2})}}-1}.
  \end{aligned}
\end{equation}
Here, \(\mathcal{N}_{L} \equiv \sup_{A} \norm{LA}_{\tr} / \norm{A}_{\tr}\), where \(\norm{\cdot}_{\tr}\) denotes the trace norm, and
\begin{equation}
  \label{eq:deviation-bcf-def}
  \Delta c(\tau,\tau') \equiv {c_{0}(\tau-\tau')-{c}(\tau,\tau')}.
\end{equation}

Crucially, the bound given in \cref{eq:observable-deviation-bound-lo} depends only on the deviation \(\Delta c(\tau,\tau')\), rather than the bath correlation function itself. Unlike perturbative results or Grönwall-type inequalities, which lead to expansions that typically converge only at short times \(\tau \lesssim 1/\delta\) irrespective of the emulation quality, here the bound remains informative as long as the cumulative (time-integrated) error is small \cite{Huang2024}.

In the specific context of the Wigner-Weisskopf model (\cref{sec:appl-cond-wign}), the excited-state probability \(p_{e}(\tau)\) serves as the primary observable. With \(O=\ketbra{e}-\ketbra{g}\), \(L=\ketbra{g}{e}\), and the normalization \(\mathcal{N}_{O}=\mathcal{N}_{L}=1\), we have $\ev{O(\tau)}=2 p_e(\tau)-1$ and so the bound becomes:
\begin{equation}
  \label{eq:pe_deviation_bound}
  \Delta p_{e}(\tau)
  \leq \frac{1}{2}\bqty{\exp(4 \delta^{2}\int_{0}^{\tau}\dd{\tau_{1}}\int_{0}^{\tau_{1}}\dd{\tau_{2}} \abs{\Delta c(\tau_{1}, \tau_{2})})-1}.
\end{equation}
This result provides a rigorous tool for estimating the quality of the fiber-loop emulation discussed in \cref{sec:mapping-to-ww}.

\subsection{Quantifying the deviation of the bath correlation function}
\label{sec:devi-bath-corr}

For the Wigner-Weisskopf dynamics described in the previous sections, we define the target model through the thermodynamic ($N\to\infty$) limit of \cref{eq:bcf-stat},
\begin{equation}\label{eq:c0definition}
c_0(\tau)=\int_{-\pi}^{\pi}\frac{\dd{k}}{2\pi} e^{-i\varepsilon_{s}(k)\tau}=\frac{1}{\pi}\int_{0}^{\omega_c}\dd{\omega} J(\omega)\eu^{-\iu \omega \tau},
\end{equation}
where $\varepsilon_{s}(k)\propto |k|^{1/(s+1)}$ is the ideal ($N_F\to\infty$) dispersion given in \cref{eq:dispersion-powerlaw}, resulting in $J(\omega)\propto\omega^s$. The physical model is instead described by the bath correlation function that generates evolution of the damping-compensated amplitude $\tilde{\alpha}(\tau)=e^{\gamma_\sys\tau/2}\alpha(\tau)$,
\begin{equation}
  \label{eq:physical_bcf}
  c(\tau,\tau') = \frac{1}{N}\sum_{k}\eu^{-\iu \bqty{\esk_{sk}\pqty{\tau -\tau'}+ {\phi}_{k}(\tau)-\phi_k(\tau')}-\frac{\Delta\gamma}{2} \abs{\tau-\tau'}}.
\end{equation}
Here, $\esk_{sk}$ gives the approximate dispersion arising from a Fourier-truncated parametric drive modulation having $\propto N_F$ Fourier coefficients, $\phi_k(\tau)$ is the nonstationary phase introduced in \cref{eq:residual-phase}, and $\Delta\gamma=\gamma_\bath-\gamma_\sys$ is the damping asymmetry. The ideal model is therefore recovered from the physical model [i.e., $\Delta c(\tau,\tau')\to 0$] in the limits $N\to\infty$, $N_F\to\infty$, $\phi_k(\tau)\to 0$, and $\Delta\gamma\to 0$. In what follows, we derive an upper bound on $\Delta c(\tau,\tau')$ depending on each of these four parameters. \Cref{eq:c0definition,eq:physical_bcf}, together with \cref{eq:pe_deviation_bound}, were used to numerically obtain the error funnels in \cref{fig:comp_ideal_v_real}.

To isolate the contribution of each non-ideal parameter step-by-step, we introduce shorthand versions of the physical correlation function \(c(\tau,\tau')\) [cf.~\cref{eq:physical_bcf}] where select non-idealities are successively removed. We let \(c(\tau,\tau') = c_{\Delta\gamma,\phi,N,N_{F}}\) be the full physical correlation function, \(c_{\phi,N,N_{F}}\) be its value for zero damping asymmetry (\(\Delta\gamma = 0\)), and \(c_{N,N_{F}}\) be the value when both \(\Delta\gamma = 0\) and the non-stationary phase vanishes (\(\phi_{k} = 0\)). This allows us to decompose the overall deviation $\Delta c=c(\tau,\tau')-c_0(\tau-\tau')$ in successive steps as:
\begin{align}
  \label{eq:triangle-ineq-c}
  \abs{\Delta c} &=
                  \begin{multlined}[t]
                    \big|c_{\Delta\gamma,\phi,N,N_{F}} - c_{\phi,N,N_{F}} + c_{\phi,N,N_{F}} - c_{N,N_{F}}\\+c_{N,N_{F}}-c_{0}\big|
                  \end{multlined}\\
                &\leq                  \begin{multlined}[t]
                  \abs{c_{\Delta\gamma,\phi,N,N_{F}} - c_{\phi,N,N_{F}}} + \abs{c_{\phi,N,N_{F}} - c_{N,N_{F}}}\\+\abs{c_{N,N_{F}}-c_{0}}.                                       \end{multlined}\label{eq:second-step}
\end{align}
We then introduce the four residuals \(c_{\Delta\gamma}(\tau),\,\Delta c_{\phi},\,\Delta c_{N_{F}}(\tau)\),  and \(\Delta c_{N}(\tau)\), satisfying the following inequalities:
\begin{align}
  \label{eq:triangle-components-delta-gamma}
  \abs{c_{\Delta\gamma,\phi,N,N_{F}} - c_{\phi,N,N_{F}}} &\leq \Delta c_{\Delta\gamma}(\tau), \\
  \label{eq:triangle-components-phi}
  \abs{c_{\phi,N,N_{F}} - c_{N,N_{F}}} & \leq \Delta c_{\phi},\\
  \label{eq:triangle-components-NF-N}
  \abs{c_{N,N_{F}}-c_{0}} & \leq \Delta c_{N_{F}}(\tau) + \Delta c_{N}(\tau).
\end{align}
The sum of these residuals then bounds \(\Delta c\):
\begin{equation}
  \label{eq:bcf_deviation_bound_sum}
  \abs{\Delta c(\tau,\tau')} \leq \begin{multlined}[t]
    \Delta c_{\Delta\gamma}(\tau-\tau') + \Delta c_{\phi}\\
    +\Delta c_{N_{F}}(\tau-\tau') + \Delta c_{N}(\tau-\tau') .
  \end{multlined}
\end{equation}
We find the following expressions for the residuals given in \cref{eq:triangle-components-delta-gamma,eq:triangle-components-phi,eq:triangle-components-NF-N} (see \cref{sec:deta-disc-conv} for details):
\begin{align}
  \label{eq:residual_delta_gamma_asymptotic}
  \Delta c_{\Delta\gamma}(\tau) &= \abs{1-\eu^{-\frac{\Delta\gamma}{2}\abs{\tau}}},\\
  \label{eq:residual_phi_bound}\Delta c_{\phi} &= \max_{k,\tau,\tau'}\abs{\phi_{k}(\tau)-\phi_{k}(\tau')}\leq2\pi\frac{\omega_{c}}{\Omega},\\
  \label{eq:residual_N_bound}{\Delta c}_{N}(\tau) &= {\frac{2\eu^{\abs{\tau}2 \frac{v_{G}}{N_{F}}}}{\eu^{\frac{N}{N_{F}}}-1}},\\
    \label{eq:residual_NF_bound}{\Delta c}_{N_{F}}(\tau) &= \tau\sqrt{\int_{-\pi}^{\pi}\frac{\dd{k}}{2\pi}\abs{\esk_{sk}-\varepsilon_{s}(k)}^{2}}.
\end{align}
Here,
\begin{equation}
  \label{eq:max_group_velocity_def}
    v_{G}\equiv\max_{k}\abs{\dv{\esk_{sk}}{k}} \lesssim\frac{\omega_{c}}{2\sqrt{\pi} s} \bqty{N_{F}^{\frac{s}{1+s}}\pqty{1+s} - 1} \leq \frac{\omega_{c}N_{F}}{2\sqrt{\pi}}
\end{equation}
is the maximal group velocity of a fictitious particle propagating in the synthetic dimension. \Cref{eq:residual_phi_bound} is valid when \(v_{l}(\tau)=v(\tau)\) is independent of \(l\) and \(N\to\infty\).

Focusing on the case \(\Delta p_{e}\ll1\), we expand the exponential in \cref{eq:pe_deviation_bound} to leading order in the coupling \(\delta\). Then, using asymptotic expressions for the individual residuals (in the limit \(N\gg N_{F}\gg1\)), we find the following parametric expression for the upper bound on \(\Delta p_{e}\):
\begin{equation}
  \label{eq:final-bound}
  \begin{aligned}
    \Delta p_{e}(\tau) \lesssim \delta^{2}\Bigg[& \frac{\omega_{c}}{12 \sqrt{\pi}}N_{F}^{-\frac{1}{2}\frac{3+s}{1+s}} {\tau^{3}}
      + \frac{\Delta\gamma}{6}\tau^{3} \\
      &+  2\pi\frac{\omega_{c}}{\Omega}{\tau^{2}}
+{\frac{2\pi}{\omega_{c}^{2}}{\eu^{-\frac{N}{N_{F}}}}\pqty{\eu^{\frac{\omega_{c}\abs{\tau}}{\sqrt{\pi}}}-\frac{\omega_{c}\abs{\tau}}{\sqrt{\pi}}-1}}
    \Bigg].
  \end{aligned}
\end{equation}
The contribution  \(\propto \omega_{c} / \Omega\) arising from the nonstationary phase typically overestimates the size of the error, because additional potential averaging of $e^{i\phi_k(\tau)}$ over $k$ and $\tau$ is neglected in deriving the bound. A tighter bound can be derived by considering further details of the full system-bath model, as discussed in \cref{sec:rotat-wave-appr}.

\section{Conclusions and outlook}
\label{sec:conclusion}
In this paper we have proposed a scheme for emulating quantum system-bath models using a parametrically driven cavity to synthesize a bath with a tunable spectral density. Because the form of the synthetic spectral density is determined solely by the parametric drive's Fourier coefficients, this form can be tuned straightforwardly with a waveform generator. In addition, we showed how one can independently verify the form of the spectral density by directly probing the cavity used to emulate the bath.
As a concrete application, we proposed an experimental setup that can emulate dynamics of the Wigner-Weisskopf model~\cite{Weisskopf1930,Scully1997} with a bath with a power-law spectral density \(J(\omega)\sim \omega^{s}\). This setup can be realized by driving classical electromagnetic field modes supported by optical fiber-loop cavities~\cite{Dutt2019,Yang2022a,Senanian2023,Pellerin2024}.

For this Wigner-Weisskopf model, we have found conditions that realize a robust transition from decay to non-decay for an atom coupled to a bath that is tuned from the subohmic ($s<1$) to the superohmic ($s>1$) regime. Conventionally, this model is analyzed in the limits of infinite bath size \(N\to\infty\) and vanishing coupling strength \(\delta\to0\). These ideal limits may not be reached in experiment and finite values may lead to a tradeoff. In particular, we have found that the number of bath modes $N$ must be exponentially large in the inverse coupling strength $1/\delta$ to realize a sharp transition at $s=1$, \(N\gg 2 \eu^{(\omega_{c} / \delta)^{2}}\) [cf.~\cref{eq:pe_ohmic}].

We have found a bound on systematic errors that may be realized due to experimental imperfections for the dynamics of any system observable $\ev{O(\tau)}$, and specifically for the excited-state probability, $p_e(\tau)$, of a decaying atom in the Wigner-Weisskopf model. In particular, we have found dependencies of this bound on damping rates, the number of parametric-drive Fourier coefficients $N_F$, the bath size $N$, the system-bath coupling strength $\delta$, and the cutoff frequency $\omega_c/2\pi$ relative to the cavity free spectral range $\Omega/2\pi$. Finally, we have shown numerically (\cref{fig:comp_ideal_v_real}) that, for parameters that have been realized in a previous experiment~\cite{Pellerin2024} on optical fiber loops, the dynamics of the ideal Wigner-Weisskopf model (\(N\to\infty\), \(N_{F}\to\infty\)) can be accurately realized in the proposed fiber-loop setup pictured in \cref{fig:new_snowman}.

The interplay between topology and dissipation has attracted significant interest recently~\cite{Rudner2009,Ozawa2019a,Yang2022,Okuma2023,Garcia-Garcia2025,Chaduteau2026,Okuma2020,Mera2026}. Many works in this area rely on non-Hermitian or Lindbladian descriptions of the interaction with the bath, which are valid only for weak coupling and in the Markovian limit. Beyond these limits, novel phenomena can arise. One such example is the non-Markovian quantum walk discussed in \refcite{Ricottone2020}, where  topological quantization of the mean displacement of a hopping particle is destroyed depending on the spectral density of the non-Markovian bath. This effect can be traced back to the behavior of the Wigner-Weisskopf model discussed above. The scheme presented in the present paper may be useful for identifying similar effects in more complex situations such as two, three, or even four dimensions~\cite{Zilberberg2018,Bouhiron2024,Lohse2018}, or with higher-order topological states~\cite{Benalcazar2017,ElHassan2019}.

As the scheme presented here allows for direct access to bath observables, it provides a potential platform for studying the flow of energy between the system and the bath, which is important for quantum thermodynamics, especially beyond the weak-coupling limit~\cite{Wiedmann2021,Kosloff2014,Boettcher2024}. A necessary step towards applications of this scheme to quantum thermodynamics would be the generalization to finite-temperature baths or to non-equilibrium bath states, such as squeezed states or even non-Gaussian bosonic states.

The bath-emulation scheme presented here could also potentially be combined with digital fault-tolerant quantum simulation of an otherwise classically intractable quantum system. Expanding a fault-tolerant simulation to the whole universe (including system and bath) may have an enormous cost in resources (gates and physical qubits). Instead, it may be possible to incorporate a non-fault-tolerant bath module, as described in this work, and to bound the error in simulating a system observable in contact with the bath, as we have done in \cref{sec:fidelity-finite-size}.

\begin{acknowledgments}
  V.B., F.P., P.S.-J., and W.A.C. acknowledge financial support from the Natural Sciences and Engineering Research Council (NSERC) and the Fonds de Recherche du Qu\'ebec-Nature et Technologies (FRQNT). We also acknowledge support from the Institut Transdisciplinaire d'Information Quantique (INTRIQ) and the Regroupement Qu\'eb\'ecois sur les Mat\'eriaux de Pointe (RQMP).
\end{acknowledgments}

\appendix
\crefalias{section}{appendix}
\crefalias{subsection}{appendix}
\crefalias{subsubsection}{appendix}
\section{Fourier expansion of the power-law dispersion relation}
\label{sec:dispersion}
The ideal Fourier coefficients, \cref{eq:hop-amp}, are given by
\begin{equation}
  \label{eq:hypergeometric-coefficients-formula}
  t_{sn}=\frac{\omega_{c}}{2}\frac{\Gamma \left(\frac{s + 2 }{2 s +2}\right)
    {}_1F_2\left(\frac{s+2 }{2 s +2};\frac{1}{2},\frac{3}{2}+\frac{1}{2 (s +1)};-\frac{1}{4} n^2 \pi
    ^2\right)}{\Gamma \left(\frac{3}{2}+\frac{1}{2 (s +1)}\right)},
\end{equation}
where \({}_{1}F_{2}\) is a generalized hypergeometric function~\cite{Olver2010}.  For \(n\gg1\) and \(s>0\), these coefficients have the asymptotic expansion~\cite{Nijimbere2018}
\begin{equation}
  \label{eq:asymptotic-decay-expansion-full}
  \begin{aligned}
    t_{sn}  & =-\frac{\omega_{c}}{2}\xi_{s} \abs{n}^{-\frac{s +2}{s +1}}+ \mathcal{O}\pqty{\frac{1}{n^{2}}},      \\
    \xi_{s} & \equiv \abs{\frac{2^{\frac{2+s}{1+s}} \pi ^{-\frac{s +3}{2 s +2}} \Gamma \left(\frac{s +2}{2 s +2}\right)}{\Gamma \left(-\frac{1}{2 s +2}\right)}},
  \end{aligned}
\end{equation}
where the coefficient \(\xi_{s}\) is bounded:
\begin{equation}
  \label{eq:2}
  \xi_{s}\leq \frac{1}{2\sqrt{\pi}}.
\end{equation}
This expansion will be used to characterize corrections to the bath correlation function due to a finite-bandwidth parametric drive in \cref{sec:finite-modul-bandw}, below.
As the dispersion is continuous and periodic [$\varepsilon_{s}(\pi) = \varepsilon_{s}(-\pi)$], the Fourier coefficients given above yield a Fourier series that converges absolutely and uniformly~\cite{Tolstov1976}.

\section{Deviations in the bath correlation function}
\label{sec:deta-disc-conv}

In this appendix, we derive the expressions for the residuals introduced in \cref{eq:residual_delta_gamma_asymptotic,eq:residual_phi_bound,eq:residual_N_bound,eq:residual_NF_bound} of the main text.

\subsection{Influence of the damping asymmetry}
\label{sec:photon-loss}

Using the definition of the physical bath correlation function in \cref{eq:physical_bcf}, we can bound the residual due to photon loss, \(\Delta c_{\Delta\gamma}\):
\begin{align}
  \label{eq:residual_delta_gamma_derivation}
  \Delta c_{\Delta \gamma}(\tau,\tau') &= \abs{c_{\Delta\gamma,\phi,N,N_{F}}(\tau,\tau')-c_{\phi,N,N_{F}}(\tau,\tau')}\\
  {}&=
    \abs{c_{\phi,N,N_{F}}(\tau,\tau')\pqty{1-\eu^{-\frac{\Delta\gamma}{2}\abs{\tau-\tau'}}}}\\
  {}&\leq \abs{1-\eu^{-\frac{\Delta\gamma}{2}\abs{\tau-\tau'}}},
\end{align}
where we have used \(\abs{c_{\phi,N,N_{F}}(\tau,\tau')}\leq1\) which follows directly from \cref{eq:physical_bcf}. This reproduces the expression given in the main text [\cref{eq:residual_delta_gamma_asymptotic}].

\subsection{Nonstationary Phase}
\label{sec:rotat-wave-appr}
\begin{figure}[t]
  \centering
  \includegraphics{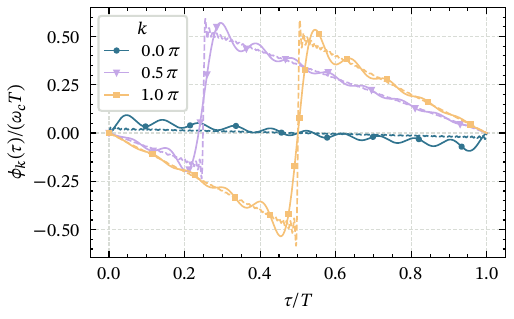}
  \caption{\label{fig:nonstat-phase} The nonstationary phase \(\phi_{k}(\tau)\) for a power-law dispersion as in \cref{eq:dispersion-powerlaw} with \(s=1\), \(N_{F}\to\infty\) and \(v_{l}(\tau)=v(\tau)\) for multiple values of \(k\) (see legend), \(M=10\) (solid lines), and \(M=100\) (dashed lines).}
\end{figure}

To isolate the effect of the nonstationary phase \(\phi_{k}(\tau)\), we compute
\begin{align}
  \label{eq:residual_phi_derivation}
  \big|c&_{\phi,N,N_{F}}(\tau,\tau') - c_{N,N_{F}}(\tau-\tau')\big|\\
    &=\abs{\frac{1}{N}\sum_{k}\eu^{-\iu \bqty{\esk_{sk}\pqty{\tau -\tau'}+ {\phi}_{k}(\tau)-\phi_k(\tau')}}-\eu^{-\iu \bqty{\esk_{sk}\pqty{\tau -\tau'}}}}\label{eq:diff-expanded}\\
         &\leq \frac{2}{N}\sum_{k}\abs{\sin\pqty{\frac{{\phi}_{k}(\tau)-\phi_k(\tau')}{2}}}\\
  &\leq     \frac{1}{N}\sum_{k}\abs{{\phi}_{k}(\tau)-\phi_k(\tau')}\label{eq:phase-discarded}\\
    &\leq \max_{\tau,\tau'}\abs{\phi_{k}(\tau)-\phi_{k}(\tau')}\label{eq:maxbound}.
\end{align}
Combining \cref{eq:residual-phase,eq:time-depdendent-dispersion} we have
\begin{equation}
  \label{eq:concrete-nonstat-phase}
  \phi_{k}(\tau)  = \int_{0}^{\tau}\bqty{\sum_{l=-M}^{M}\eu^{-\iu l \pqty{k-\Omega \tau' }}v_{l}(\tau')-\esk_{sk}}\dd{\tau'}.
\end{equation}
If the periodic drive amplitude \(v_l(\tau)\) is approximately independent of \(l\) for \(l\lesssim N_F\) and assuming \(M = (N-1) / 2 \gg N_{F}\), then we can approximate the above for \(0\leq\tau\leq T\) and \(0\leq k \leq 2\pi\) by setting \(v_{l}(\tau) = v(\tau)\) for all $l$ and taking $M\to\infty$:
\begin{equation}
  \label{eq:simplified-nonstat}
  \begin{aligned}
    \phi_{k}(\tau) & = \int_{0}^{\tau}\bqty{v(\tau')\sum_{l=-\infty}^{\infty}\eu^{-\iu l \pqty{k-\Omega \tau'}}-\esk_{sk}}\dd{\tau'} \\
                   & = v({k} / {\Omega})T\Theta(\Omega\tau -k) - \esk_{sk}\tau                                           \\
                   & = \bqty{T\Theta(\Omega\tau -k)-\tau}\esk_{sk},
  \end{aligned}
\end{equation}
where we used \(v(k / \Omega) = \esk_{sk}\) [see \cref{eq:epsilon-approximate-dispersion}]. As \(\phi_{k}(\tau)\) is periodic with period \(T= 2\pi / \Omega\), \cref{eq:simplified-nonstat} describes a sawtooth pattern (\cref{fig:nonstat-phase}). We therefore have
\begin{equation}
  \label{eq:finalphi}
  \max_{\tau,\tau'}\abs{\phi_{k}(\tau)-\phi_{k}(\tau')} \leq \esk_{sk} {T} \leq \omega_{c}T=2\pi \frac{\omega_{c}}{\Omega }.
\end{equation}
This reproduces \cref{eq:residual_phi_bound} in the main text.

The bound given in \cref{eq:maxbound} will often be overly pessimistic. In particular, the step going from \cref{eq:diff-expanded} to \cref{eq:phase-discarded} neglects the possibility for rapid phase averaging with \(k\). In addition, phase oscillations with \(\tau\) can lead to further averaging when integrating the equations of motion. We therefore typically expect better agreement between the physical and target model than is implied by \cref{eq:maxbound,eq:finalphi}.

\subsubsection{Averaging out rapid oscillations}
\label{sec:averaging-out-rapid}
In this appendix we compute corrections to the spectral density \(J(\omega)\) due to the presence of the nonstationary phase \(\phi_{k}(t)\). To obtain an effective equation-of-motion for the part of an observable [e.g., the excited-state amplitude $\alpha(\tau)$] that varies slowly on the timescale $T=2\pi/\Omega$, we consider an effective bath correlation function that is coarse-grained in time:
\begin{align}
  \label{eq:coarse-grain-def}
  \tilde{c}(\tau,\tau')= \frac{1}{T^{2}} \int_{-\frac{T}{2}}^{\frac{T}{2}}\dd{\mu}\int_{-\frac{T}{2}}^{\frac{T}{2}}\dd{\mu'}c(\tau+\mu,\tau'+\mu'),
\end{align}
where \(c(\tau,\tau')\) is the bath correlation function including the nonstationary phase, given in \cref{eq:physical_bcf}. Under the conditions $\omega_a^*\ll\Omega$, $\Delta\gamma\ll \Omega$, $\omega_c\ll \Omega$, all terms evolving with the associated scales ($\omega_a^*,\Delta\gamma,\omega_c$) can be taken to be approximately constant over the period $T$, leading to an approximate time-local effective bath correlation function $\tilde{c}(\tau,\tau')\simeq \tilde{c}(\tau-\tau')$:
\begin{align}
  \label{eq:final-coarse}
   \tilde{c}(\tau) &=\frac{1}{N}\sum_{k}\eu^{-\iu {\esk_{sk}\tau}-\frac{\Delta\gamma}{2} \abs{\tau}}\sinc^{2}\pqty{\pi\frac{{\esk_{sk}}}{\Omega}},
\end{align}
where we have assumed that the drive amplitude \(v_{l}(\tau) = v(\tau)\) is approximately independent of \(l\) for \(l\lesssim N_F\) and we have used the expression for $\phi_k(\tau)$ given in \cref{eq:simplified-nonstat}, which is valid in the limit \(N\to\infty\). In the limit $\Delta\gamma\to 0$, Fourier transforming \cref{eq:final-coarse} leads to a coarse-grained spectral density:
\begin{align}
  \label{eq:new-sd}
  \tilde{J}(\omega) &= J(\omega) \sinc^{2}\pqty{\pi\frac{\omega}{\Omega}}.
\end{align}
Thus, since $\sinc\left(\pi\omega/\Omega\right)\simeq 1-\mathcal{O}\left[\left(\omega/\Omega\right)^2\right]$, observables that depend on the low-frequency ($\omega\to 0$) behavior of $J(\omega)$, such as the long-time average $\peb$, are reproduced accurately.

Finally, the system-bath model explored in the main text considers coupling only to a single mode labeled with the index $n=0$. This can be justified either in the case where the physical coupling to higher modes vanishes or within a strict rotating-wave approximation. To account for corrections beyond the rotating-wave approximation, we could consider the more general form:
\begin{equation}
  \label{eq:coupleall}
  \widetilde{H}_{I} \to \widetilde{H}_{I}'=\sum_{n=-M}^{M}\pqty{\delta_{n}a^{\dag}{b}_{n}+\hc}.
\end{equation}
Specifically, for the fiber-loop setup of \cref{sec:appl-cond-wign}, applying coupled-mode theory~\cite{Snyder1972} to describe a pair of optical fibers coupled over distance $d$ gives
\begin{equation}
  \label{eq:9}
  \delta_{n} = \delta\sinc \frac{\pi n}{M_{c}}= \delta\sinc{\frac{{d n\Omega n_{0}}}{c}},
\end{equation}
where $M_c=\pi c/d\Omega n_0$, \(n_{0}\approx 1.5\) is the refractive index, and \(c\) is the speed of light in vacuum. For \(d=\SI{5}{\centi\meter}\) and \(\Omega/2\pi=\SI{10}{\mega\hertz}\) we find \(M_{c}\approx 200\).

We rederive the two-time correlation function $c(\tau,\tau')$ using the modified system-bath coupling $\widetilde{H}_{I}'$, then perform the same coarse-graining procedure described before \cref{eq:final-coarse}. When the system-bath coupling parameters $g_n$ are given by \cref{eq:9}, we then find the coarse-grained time-local bath correlation function
\begin{equation}
  \label{eq:5}
  \tilde{c}(\tau) = \frac{1}{N}\sum_{k}\eu^{-\iu {\esk_{sk}\tau}-\frac{\Delta\gamma}{2} \abs{\tau}} \sinc^{2}\pqty{\frac{\pi \esk_{sk}}{\Omega M_{c}}}.
\end{equation}
Notably, when the correct form of $\delta_n$ is taken into account for coupled fiber loops, we find an even stronger suppression of the corrections to the rotating-wave approximation, of order $(\pi\esk_{sk}/\Omega M_c)^2\ll (\pi\esk_{sk}/\Omega)^2$ for $M_c\gg 1$. This additional suppression is due to stronger phase averaging from additional contributions $\propto \delta_n$ for $n\ne 0$.

\subsubsection{Simulations beyond the rotating-wave approximation}
\label{sec:rotat-wave-appr-1}
In this appendix we show numerically that the dynamics of \(p_{e}(\tau)\eu^{\gamma \tau}\) are well reproduced within the rotating-wave approximation when \(\omega_{c} / \Omega\ll1\).
To achieve this, we integrate a modified version of the equations of motion [\cref{eq:coupled-amplitude-evolution}], but where the system-bath coupling is given by the more general form in \cref{eq:coupleall} with $\delta_n= \delta$ for all relevant $n$, and including counter-rotating terms that oscillate at frequencies up to \(n^{*}\Omega\), where we chose \(n^{*}=10\):
\begin{equation}
  \label{eq:coupled-amplitude-evolution-generalized}
  \begin{aligned}
    \dot{\alpha}    & = -\iu \omega_{a}\alpha  -\frac{\gamma}{2} \alpha -\iu \frac{\delta}{\sqrt{N}} \sum_{k} \sum_{\abs{n}\leq n^{*}}\eu^{-\iu(\Omega n \tau - kn)}\beta_{k},             \\
    \dot{\beta}_{k} & = -\iu \varepsilon^{n^{*}}_{k}(\tau) \beta_{k} - \frac{\gamma}{2} \beta_{k} -\iu \frac{\delta}{\sqrt{N}} \sum_{\abs{n}<{n^{*}}}\eu^{\iu(\Omega n \tau - kn)} \alpha,
  \end{aligned}
\end{equation}
where
\begin{equation}
  \label{eq:time-dep-disp-sim}
  \varepsilon^{n^{*}}_{k}(\tau) = \sum_{j=-N_{F}}^{N_{F}}\sum_{\abs{l-j}<n^{*}}t_{sj}\eu^{-i lk} \eu^{-i (j-l)\Omega\tau}
\end{equation}
is a Fourier-truncated version of the time-dependent dispersion relation \cref{eq:time-depdendent-dispersion} with the \(t_{sj}\) being the Fourier coefficients given in \cref{eq:hypergeometric-coefficients-formula} of the power-law dispersion relation \cref{eq:dispersion-powerlaw}. In \cref{eq:time-dep-disp-sim}, we have assumed that the parametric drive amplitudes \(v^{j}_{l}=t_{sj}\) are independent of the bath-cavity mode frequency difference \(l\Omega\).

\begin{figure}[t]
  \centering
 \includegraphics{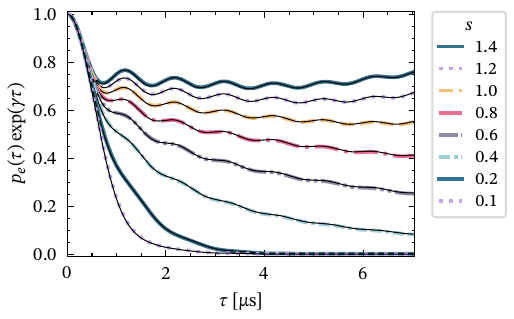}
  \caption{\label{fig:wo_rwa} Comparison between the realistic model including counter-rotating terms oscillating at frequencies up to \(n^*\Omega\) with \(n^*=10\) [colored lines, resulting from integrating \cref{eq:coupled-amplitude-evolution-generalized}] and without counter-rotating terms (thin black lines, corresponding to the colored lines in \cref{fig:comp_ideal_v_real}). For the experimental parameters in \cref{tab:simuparam} (\(\omega_c/\Omega \approx 0.1\)), the non-stationary phase corrections are negligible.}
\end{figure}
\begin{figure}[t]
  \centering
  \includegraphics{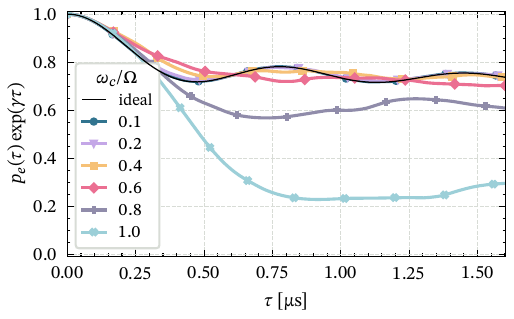}
  \caption{\label{fig:rwa_breakdown} Excited-state probability \(p_{e}(\tau)\) [\cref{eq:excited-state-population-probability}] obtained by numerically integrating \cref{eq:coupled-amplitude-evolution-generalized} including counter-rotating terms up to \(n^{*}=10\) for a superohmic bath (\(s=1.5\)) at various ratios \(\omega_{c}/\Omega\) (colored lines). The black dashed line indicates the ideal limit (\(\omega_{c} / \Omega \to 0\)). For \(\omega_{c} / \Omega \lesssim 0.2\), the rotating-wave approximation remains accurate, but non-stationary corrections become significant at higher ratios. Other parameters match those in \cref{fig:comp_ideal_v_real} with \(N_{F}=500\).}
\end{figure}

The pessimistic error bound that was given in \cref{eq:final-bound} would suggest a possible relative error of \(100\%\) at \(\tau\approx\SI{1}{\micro\second}\) for the parameters in \cref{tab:simuparam} (with \(\omega_{c} / \Omega\approx 0.1\)). However, as shown in \cref{fig:wo_rwa}, we find that the nonstationary corrections are much smaller than $p_e(\tau)$ itself up to $\tau\gtrsim 6\,\mu\mathrm{s}$ using the same parameters. \Cref{fig:rwa_breakdown} demonstrates that the rotating-wave approximation breaks down for \(\omega_{c} / \Omega\gtrsim 1\) as expected.

\subsection{Drive bandwidth}
\label{sec:finite-modul-bandw}
In this appendix we quantify the effect of the finite number of drive Fourier coefficients \(\propto N_{F}\). We aim to bound
\begin{align}
  \label{eq:bandwidth_error_split}
  \abs{c_{N,N_{F}}(\tau)-c_{0}(\tau)}
  &= \abs{\frac{1}{N}\sum_{k}\eu^{-\iu \esk_{sk}\tau} - \int_{-\pi}^{\pi}\eu^{-\iu \varepsilon_{s}(k) \tau}\frac{\dd{k}}{2\pi}} \tag{B13} \\
  &\le \abs{\frac{1}{N}\sum_{k}\eu^{-\iu \esk_{sk}\tau} - \int_{-\pi}^{\pi}\eu^{-\iu\esk_{sk}\tau}\frac{\dd{k}}{2\pi}} \nonumber \\
  &\quad + \abs{\int_{-\pi}^{\pi}\eu^{-\iu\esk_{sk}\tau}-\eu^{-\iu \varepsilon_{s}(k) \tau}\frac{\dd{k}}{2\pi}}\label{eq:added-zero}  \\
  &\le {\Delta c_{N}(\tau)} + \abs{\int_{-\pi}^{\pi} \left( \eu^{-\iu \esk_{sk}\tau} - \eu^{-\iu\varepsilon_{s}(k) \tau} \right) \frac{\dd{k}}{2\pi}} \label{eq:discret-error}
\end{align}
where \(\esk_{sk}\) is the Fourier-truncated dispersion [\cref{eq:filtered_bath_energies}] and \(\varepsilon_{s}(k)\) is the ideal dispersion relation defined in \cref{eq:dispersion-powerlaw}. To go from \cref{eq:added-zero} to \cref{eq:discret-error}, the sum over $k$ is changed to an integral, resulting in an upper bound on the remainder term \(\Delta c_{N}(\tau)\) [given in \cref{eq:residual_N_bound}]. The explicit steps are explained in \cref{sec:residual}, below.

We now proceed to bound the second term on the right-hand side of \cref{eq:discret-error},
\begin{align}
  \label{eq:whattobound}
  \Bigg|\int_{-\pi}^{\pi}\eu^{-\iu \esk_{sk}\tau}& - \eu^{-\iu \varepsilon_{s}(k) \tau}\frac{\dd{k}}{2\pi}\Bigg| \\
                                          &\leq  {\int_{-\pi}^{\pi}\abs{\eu^{-\iu \esk_{sk}\tau} - \eu^{-\iu \varepsilon_{s}(k) \tau}}\frac{\dd{k}}{2\pi}}\\
  \label{eq:cs} &\leq\sqrt{{\int_{-\pi}^{\pi}\abs{\eu^{-\iu \esk_{sk}\tau} - \eu^{-\iu \varepsilon_{s}(k) \tau}}^{2}\frac{\dd{k}}{2\pi}}}\\
                                                                                               \label{eq:meanval}&\leq \tau\sqrt{\int_{-\pi}^{\pi}\abs{\esk_{sk} - \varepsilon_{s}(k)}^{2}\frac{\dd{k}}{2\pi}}\\
                                                                                                &= \Delta c_{N_{F}}(\tau),
\end{align}
where we have used the Cauchy-Schwartz inequality in \cref{eq:cs} and the mean-value theorem in \cref{eq:meanval}.  This reproduces the expression given in \cref{eq:residual_NF_bound}.

To evaluate \(\Delta c_{N_{F}}(\tau)\) we apply Parseval's theorem to the integral in \cref{eq:meanval}:
\begin{equation}
  \label{eq:3}
  {\int_{-\pi}^{\pi}\abs{\esk_{sk} - \varepsilon_{s}(k)}^{2}\frac{\dd{k}}{2\pi}} = 2\sum_{n=N_{F}+1}^{\infty}\abs{t_{sn}}^{2},
\end{equation}
where we have used the fact that \(\esk_{sk}\) [\cref{eq:filtered_bath_energies}] is the truncated Fourier series of \(\varepsilon_{s}(k)\) and that \(\abs{t_{sn}}=\abs{t_{s,-n}}\). Using the asymptotic expansion of the Fourier coefficients \(t_{sn}\) given in \cref{eq:asymptotic-decay-expansion-full}
and the Euler--Maclaurin formula, we then find to leading-order in \(N_{F}\gg1\):
\begin{equation}
  \label{eq:final-asymp}
  \Delta c_{N_{F}}(\tau)\approx \omega_{c} \tau \xi_{s} \sqrt{\frac{1+s}{4 (3+s)}}N_{F}^{-\frac{1}{2}\frac{3+s}{1+s}} \leq \frac{\omega_{c}\tau}{4\sqrt{\pi}}N_{F}^{-\frac{1}{2}\frac{3+s}{1+s}},
\end{equation}
where \(\xi_{s}\) is given in \cref{eq:asymptotic-decay-expansion-full}.
The final expression in \cref{eq:final-asymp} was used to obtain the first term in the square brackets in \cref{eq:final-bound} [the upper bound on \(\Delta p_{e}(\tau)\)]. Interestingly, the scaling of the residual with \(N_{F}\) switches from super-linear to sub-linear at \(s=1\).

\subsection{Number of available modes in the bath cavity}
\label{sec:numb-availb-modes}
In \cref{sec:from-long-range}, we assumed that the bath cavity has a large but finite number of accessible evenly spaced modes \(M = (N - 1) / 2\) to either side (in frequency) of the center frequency \(\omega_{0}\). The number \(M\) might be determined by the material properties of the cavity (such as mode dispersion) or wave-guide cutoff frequencies.

\subsubsection{Boundary Conditions}
\label{sec:boundary-conditions}
We found in \cref{sec:from-long-range} that the parametrically driven bath cavity realizes a long-range hopping model on a one dimensional chain. Each site in this chain corresponds to a frequency mode of the cavity. The system is coupled directly to the mode with lab-frame frequency \(\omega_{0}\) which has the site index \(n=0\). In order to assume periodic boundary conditions, it was argued that the modes far from \(n=0\) do not receive appreciable excitation on the timescale of interest. We now determine the timescale \(\tau_{\mathrm{BC}}\) at which the system dynamics become sensitive to the physical boundaries of the cavity.

An estimate of \(\tau_{\text{BC}}\) can be made by assuming that the earliest time at which the system will know of the boundaries is the time at which a pulse put into the bath at \(\tau=0\) has time to travel back and forth between \(n=0\) and \(n=M\).
The group velocity of such a pulse is bounded by
\begin{equation}
  \label{eq:max-group-velocity}
  \begin{multlined}
    v_{G}\equiv\max_{k}\abs{\dv{\esk_{sk}}{k}} = \max_{k}\abs{\sum_{n=-N_{F}}^{N_{F}}n t_{sn} \eu^{\iu n k}} \\
    \leq \sum_{n=-N_{F}}^{N_{F}} \abs{n t_{sn}} \equiv \hat{v}_{G},
  \end{multlined}
\end{equation}
with the Fourier coefficients \(t_{sn}\) as defined in \cref{eq:hypergeometric-coefficients-formula}.
Therefore, a conservative estimate of the timescale at which the boundary conditions become relevant is
\begin{equation}
  \label{eq:boundary-time-scale}
  \tau_{\mathrm{BC}}\equiv \frac{N}{\hat{v}_{{G}}} .
\end{equation}

For the power-law spectral densitiy $J(\omega)\propto \omega^{s}$ with spectral exponent \(s\), an asymptotic expression can be derived by evaluating the sum in \cref{eq:max-group-velocity} with the asymptotic expansion of the \(t_{sn}\) in \cref{eq:asymptotic-decay-expansion-full} for \(N_{F}\gg1\):
\begin{equation}
  \label{eq:asymptotic-group-speed}
  \hat{v}_{G}  \approx \frac{1}{2\sqrt{\pi} s} \bqty{N_{F}^{\frac{s}{1+s}}\pqty{1+s} - 1}.
\end{equation}

\subsubsection{Riemann-sum residual}
\label{sec:residual}
We now discuss the magnitude of the deviation \(\Delta c_{N}\) due to the approximation of the sum over \(k\) as in integral in going from \cref{eq:added-zero} to \cref{eq:discret-error}. Concretely, we aim to bound:
\begin{equation}
  \label{eq:what-to-bound-N}
  \abs{\frac{1}{N}\sum_{k}\eu^{-\iu \esk_{sk}\tau}-\int_{-\pi}^{\pi}\frac{\dd{k}}{2\pi} \eu^{-\iu \tau\esk_{sk}}}
\end{equation}
using Theorem 9.28 of \refcite{Kress2012} which bounds the error of the Riemann sum for smooth periodic function.

The energy \(\esk_{sk}\) as a function of \(k\) is periodic and holomorphic when considered on the complex plane. For
\begin{equation}
  \label{eq:complex-holomorphic-bounds}
  z= k+\iu y
\end{equation}
with \(0\leq k\leq 2\pi\), \(-N_{F}^{-1}\leq y \leq N_{F}^{-1}\), we can bound:
\begin{equation}
  \label{eq:func-upper-bound}
  \abs{\eu^{-\iu \tau \esk_{s,k+\iu y}}}\leq \eu^{\tau \epsilon_{I}},
\end{equation}
where
\begin{equation}
  \label{eq:energy-maximum-calculation}
  \epsilon_{I}\equiv 2\max_{k}\abs{\sum_{n=1}^{N_{F}}\abs{t_{sn}}\sin(kn+\varphi_{sn})\sinh(\frac{n}{N_{F}})},
\end{equation}
with \(t_{sn} = \abs{t_{sn}}\eu^{\iu \varphi_{sn}}\). Using \(\sinh(x)\leq\sinh(1)x<2x\) for \(0\leq x\leq1\), \(\abs{\sin x}\leq1\) and  \(t_{s,-n}=t^{*}_{sn}\) we can further bound \cref{eq:energy-maximum-calculation}
\begin{equation}
  \label{eq:hyperbolic-bound-velocity}
  \begin{aligned}
    \epsilon_{I} & \leq{\sinh(1) \sum_{n=-N_{F}}^{N_{F}}\abs{t_{sn}\frac{n}{N_{F}}}}
    \\
                 & = \frac{\hat{v}_{G}\sinh(1)}{N_{F}} < \frac{2 \hat{v}_{G}}{N_{F}},
  \end{aligned}
\end{equation}
where we have used \cref{eq:max-group-velocity}.

According to Theorem 9.28 of \refcite{Kress2012} (in the notation of \refcite{Kress2012}, \(M = \eu^{\tau\epsilon_{I}}\), \(a={1} / {N_{F}}\), \(n=N\), and accounting for the fact that our integration measure is \(\dd{k} / 2\pi\)), we can then obtain
\begin{equation}
  \label{eq:spectrum-fourier-dispersion}
 \abs{\frac{1}{N}\sum_{k}\eu^{-\iu \esk_{sk}\tau}-\int_{-\pi}^{\pi}\frac{\dd{k}}{2\pi} \eu^{-\iu \tau\esk_{sk}}}
  \leq \frac{2\eu^{\tau \epsilon_{I}}}{\eu^{\frac{N}{N_{F}}} -1}.
\end{equation}
Using the bound for \(\epsilon_{I}\) [\cref{eq:hyperbolic-bound-velocity}] in \cref{eq:spectrum-fourier-dispersion} yields the residual given in the main text [\cref{eq:residual_N_bound}]:
\begin{equation}
  \label{eq:residual-n-in-app}
  {\Delta c}_{N}(\tau) = {\frac{2\eu^{\abs{\tau}2 \frac{\hat{v}_{G}}{N_{F}}}}{\eu^{\frac{N}{N_{F}}}-1}}.
\end{equation}

\section{Long-time dynamics of the Wigner-Weisskopf model}
\label{sec:deta-wign-weissk}
\label{sec:some-notes-therm}
In this appendix, we justify the expressions given for the time-averaged excited-state probability \(\peb\) [\cref{eq:p_e_avg,eq:weak_coupling_suppression}] and the leading long-time dynamical behavior of the excited-state probability \(p_{e}(\tau)\) [\cref{eq:aysmp-fit}].  To this end, we analyze the solution to the equation of motion for the excited-state amplitude \(\alpha(\tau)\) [\cref{eq:excited-state-dynamics-int}], as given by its Laplace transform in \cref{eq:laptransformed-solution}:
\begin{equation}
  \label{eq:13}
    \alpha(z) = \frac{\alpha_0}{z+i\omega_a+\delta^2 c(z)}.
  \end{equation}
In the continuum limit (\(N, N_{F} \to \infty\)), the Laplace transform of the bath correlation function is
\begin{equation}
  \label{eq:czlap}
  c(z)=\frac{1}{z} {}_{2}F_{1}\pqty{1,1+s,2+s,-\frac{\iu}{z}},
\end{equation}
where \({}_{2}F_{1}\) is the Gauss hypergeometric function.

We choose the branch of the hypergeometric function so that \(\alpha(z)\) has
a branch cut along the imaginary axis between \(z=-\iu \omega_{c}\) and \(z=0\). Furthermore, there are two isolated poles at \(z=-\iu \omega_{a}^{*}\) and \(z=-\iu\hat{\omega}_{a}\) above and below the branch cut, respectively, when \(\omega_{a}^{*}<0 <\omega_{c}<\hat{\omega}_{a}\). Inverting the Laplace transform yields a time-domain solution of the form:
\begin{equation}
  \alpha(\tau) =  \overline{\alpha} \eu^{-\iu \omega_{a}^{*}\tau} + \hat{\alpha}\eu^{-\iu \hat{\omega}_{a}\tau}+\alpha_{\text{cut}}(\tau),
\end{equation}
where \(\overline{\alpha}\) and \(\hat{\alpha}\) are the residue amplitudes associated with the poles at \(z=-\iu \omega_{a}^{*}\) and \(z=-\iu \hat{\omega}_{a}\), respectively. The contribution \(\alpha_{\text{cut}}(\tau)\) arises from the branch cut.

We will now determine the approximate value of \(\hat{\omega}_{a}\) in the weak-coupling regime. For the rest of this appendix, we set \(\omega_{c}=1\). We expand the denominator of the expression for \(\alpha(z)\) given in \cref{eq:13} around the branch point by parameterizing
\begin{equation}
  \label{eq:1}
  \hat{\omega}_{a} = 1+\eu^{-\Lambda}
\end{equation}
where we assume \(\Lambda\gg1\) and neglecting all exponentially small terms:
\begin{equation}
  \label{eq:pole-expanding}
  \alpha\pqty{z=-\iu \bqty{1+\eu^{-\Lambda}}}\approx \frac{\iu}{1-\omega_{a}+(1+s)\delta^{2} \pqty{H_{s} -\Lambda}}.
\end{equation}
Above, we have used the harmonic numbers:
\begin{equation}
  \label{eq:26}
  H_{s} = \gamma_{e} +\frac{\Gamma'(1+s)}{\Gamma(1+s)}>0,
\end{equation}
which are defined in terms of the Euler-Mascheroni constant \(\gamma_{e}\approx 0.57721\) and the gamma function \(\Gamma(z)\). Solving for the zero of the denominator of \cref{eq:pole-expanding}, we find
\begin{equation}
  \label{eq:pole_location_approx}
  \Lambda = \frac{1-\omega_{a}}{(1+s)\delta^{2}}+H_{s}.
\end{equation}
We now set \(\omega^{*}_{a}=0\) by choosing
\begin{equation}
  \label{eq:4}
  \omega_{a} = \delta^{2} \frac{s}{1+s}
\end{equation}
to compensate for the Lamb shift [\cref{eq:lam}] and find
\begin{align}
  \label{eq:22}
   \Lambda &= \frac{1}{(1+s)\delta^{2}}+H_{s}-\frac{1}{s}.
\end{align}
Recall that we have assumed \(\Lambda\gg1\), which in turn constrains the coupling strength \(\delta\) for any given value of \(s\) [cf. the discussion before \cref{eq:weak_coupling_suppression}]. The corresponding residue amplitude \(\hat{\alpha}\) is:
\begin{equation}
  \label{eq:pole_residual}
  \hat{\alpha}=\frac{\alpha_{0}(\hat{\omega}_{a}-1)}{(1+s)\delta^{2}}\propto \eu^{-\Lambda},
\end{equation}
which also vanishes exponentially when \(\Lambda\gg1\). This result confirms the expression given in  \cref{eq:p_e_avg} of the main text, where we neglect contributions from the pole at \(\hat{\omega}_{a}\) to the time-averaged excited-state probability \(\peb\).

We now determine the long-time asymptotic behavior of \(\alpha_{\text{cut}}(\tau)\). The branch-cut integral is given by
\begin{equation}
  \label{eq:alphacut-def}
  \alpha_{\text{cut}}(\tau)= \int_{0}^{1}\dd{\omega} \eu^{-\iu \omega \tau} F(\omega), \quad F(\omega) \equiv \frac{\delta^{2}s^{2} (1+s)}{\omega^{s} f_{1}(\omega) f_2(\omega)},
\end{equation}
where we have introduced
\begin{align}
  f_1(\omega) & \equiv
  s \omega -i \delta ^2 (s+1) \left[ \omega ^s \left(2 \pi -i
  B_{\frac{1}{\omega }}(s+1,0)\right)-i\right]                                                 , \\
  f_2(\omega) & \equiv\delta ^2 (s+1) \left(s \omega ^s B_{\frac{1}{\omega }}(s+1,0)+1\right)-s
  \omega,
\end{align}
and \(B_{z}(a,b)\) is the incomplete beta function.
Focusing on the subohmic case (\(s < 1\)), the integrand \(F(\omega)\) features a singularity at \(\omega = 0\) that governs the long-time decay of \(\alpha_{\text{cut}}\). Splitting off the singularity of \(F(\omega)\) at \(\omega \to 0\) gives:
\begin{equation}
  \label{eq:fomegasingular}
  F(\omega) = F_{\text{rest}}(\omega)-\frac{\sin ^2(\pi  s)}{\omega^{s} \pi ^2 \delta ^2 (s+1)},
\end{equation}
where \(F_{\text{rest}}(\omega)\) is bounded and continuous for \(0 \leq \omega \leq 1\). \(F_{\text{rest}}(\omega)\) on its own would lead to a contribution to \(\alpha_{\text{cut}}\) which decays as \(\tau^{-1}\) or faster.
Evaluating \cref{eq:alphacut-def}, the singular part in \cref{eq:fomegasingular} dominates, and we find the asymptotic behavior:
\begin{equation}
  \label{eq:cutsubohm}
  \alpha_{\text{cut}}(\tau) \overset{\tau\to\infty}{\sim}
\frac{i \eu^{\frac{\iu \pi  s}{2}} \sin^2(\pi  s) \Gamma (1-s)}{\pi ^2
   \delta ^2 (s+1)} \frac{1}{\tau^{\abs{s-1}}};\quad (s<1).
\end{equation}
A similar argument can be made for the case \(1<s<2\), where we find:
\begin{equation}
  \label{eq:cutsupohm}
  \alpha_{\text{cut}}(\tau) \overset{\tau\to\infty}{\sim}
  -\frac{\iu \delta ^2 e^{-\frac{1}{2} i \pi  s} \left(s^2-1\right)
   \Gamma (s)}{\left(\delta ^2 (s+1)+s-1\right)^2}\frac{1}{\tau^{\abs{s-1}}}
  ;\quad (1<s<2).
\end{equation}
For \(s>2\), \(\alpha_{\text{cut}}(\tau)\) decays as \(\tau^{-1}\) or faster. \Cref{eq:cutsubohm,eq:cutsupohm} reproduces the for assumed for \(\alpha(\tau)\)
\begin{equation}
  e^{\gamma_S\tau}p_e(\tau)\simeq \peb+A/\tau^{|s-1|};\quad (\tau\to\infty).
\end{equation}
in \cref{eq:aysmp-fit}.

\bibliography{references} 
\end{document}